\documentstyle{mn}
\input psfig
\def\be{\begin{equation}}
\def\ee{\end{equation}}
\def\beq{\begin{eqnarray}}
\def\eeq{\end{eqnarray}}
\def\half{{\textstyle{1\over2}}}
\def\ffrac#1#2{{\textstyle\frac{#1}{#2}}}
\def\Lfour{{L_4}}
\def\Lfive{{L_5}}
\def\Lthree{{L_3}}

\def\halfi{{\textstyle{i\over2}}}
\def\aext{a_{\rm ext}}
\def\aplanet{a_{\rm p}}
\def\eplanet{e_{\rm p}}
\def\Pplanet{P_{\rm p}}
\def\nplanet{n_{\rm p}}
\def\mplanet{m_{\rm p}}
\def\rplanet{r_{\rm p}}
\def\nuplanet{\nu_{\rm p}}
\def\lambdaplanet{\lambda_{\rm p}}
\def\Mplanet{M_{\rm p}}
\def\asin{{\rm asin}}
\def\acos{{\rm acos}}
\def\yrs{\;{\rm yrs}}
\def\spose#1{\hbox to 0pt{#1\hss}}
\def\lta{\mathrel{\spose{\lower 3pt\hbox{$\sim$}}
    \raise 2.0pt\hbox{$<$}}}
\def\gta{\mathrel{\spose{\lower 3pt\hbox{$\sim$}}
    \raise 2.0pt\hbox{$>$}}}
\def\emax{e_{\rm max}}
\def\imax{i_{\rm max}}
\def\avvarpidot{\langle {\dot \varpi} \rangle}
\def\avOmegadot{\langle {\dot \Omega} \rangle}
\def\Omegadot{\dot \Omega}
\def\omegadot{\dot \omega}
\def\ijup{i_{\rm jup}}

\newif\ifAMStwofonts

\ifoldfss
  \ifCUPmtlplainloaded \else
    \NewTextAlphabet{textbfit} {cmbxti10} {}
    \NewTextAlphabet{textbfss} {cmssbx10} {}
    \NewMathAlphabet{mathbfit} {cmbxti10} {} 
    \NewMathAlphabet{mathbfss} {cmssbx10} {} 
  \fi
  \ifAMStwofonts
    \ifCUPmtlplainloaded \else
      \NewSymbolFont{upmath} {eurm10}
      \NewSymbolFont{AMSa} {msam10}
      \NewMathSymbol{\upi}     {0}{upmath}{19}
      \NewMathSymbol{\umu}     {0}{upmath}{16}
      \NewMathSymbol{\upartial}{0}{upmath}{40}
      \NewMathSymbol{\leqslant}{3}{AMSa}{36}
      \NewMathSymbol{\geqslant}{3}{AMSa}{3E}

      \let\leq=\leqslant 
       
    \fi
  \fi
\fi 

\ifnfssone
  \newmathalphabet{\mathit}
  \addtoversion{normal}{\mathit}{cmr}{m}{it}
  \addtoversion{bold}{\mathit}{cmr}{bx}{it}
  \newmathalphabet{\mathbfit} 
  \addtoversion{normal}{\mathbfit}{cmr}{bx}{it}
  \addtoversion{bold}{\mathbfit}{cmr}{bx}{it}
  \newmathalphabet{\mathbfss} 
  \addtoversion{normal}{\mathbfss}{cmss}{bx}{n}
  \addtoversion{bold}{\mathbfss}{cmss}{bx}{n}
  \ifAMStwofonts
    \ifCUPmtlplainloaded \else
      \UseAMStwoboldmath
      \makeatletter
      \new@mathgroup\upmath@group
      \define@mathgroup\mv@normal\upmath@group{eur}{m}{n}
      \define@mathgroup\mv@bold\upmath@group{eur}{b}{n}
      \edef\UPM{\hexnumber\upmath@group}
      \new@mathgroup\amsa@group
      \define@mathgroup\mv@normal\amsa@group{msa}{m}{n}
      \define@mathgroup\mv@bold\amsa@group{msa}{m}{n}
      \edef\AMSa{\hexnumber\amsa@group}
      \makeatother
      \mathchardef\upi="0\UPM19
      \mathchardef\umu="0\UPM16
      \mathchardef\upartial="0\UPM40
      \mathchardef\leqslant="3\AMSa36
      \mathchardef\geqslant="3\AMSa3E

      \let\leq=\leqslant 

    \fi
  \fi
\fi 

\ifnfsstwo
  \DeclareMathAlphabet{\mathbfit}{OT1}{cmr}{bx}{it}
  \SetMathAlphabet\mathbfit{bold}{OT1}{cmr}{bx}{it}
  \DeclareMathAlphabet{\mathbfss}{OT1}{cmss}{bx}{n}
  \SetMathAlphabet\mathbfss{bold}{OT1}{cmss}{bx}{n}
  \ifAMStwofonts
    \ifCUPmtlplainloaded \else
      \DeclareSymbolFont{UPM}{U}{eur}{m}{n}
      \SetSymbolFont{UPM}{bold}{U}{eur}{b}{n}
      \DeclareSymbolFont{AMSa}{U}{msa}{m}{n}
      \DeclareMathSymbol{\upi}{0}{UPM}{"19}
      \DeclareMathSymbol{\umu}{0}{UPM}{"16}
      \DeclareMathSymbol{\upartial}{0}{UPM}{"40}
      \DeclareMathSymbol{\leqslant}{3}{AMSa}{"36}
      \DeclareMathSymbol{\geqslant}{3}{AMSa}{"3E}

      \let\leq=\leqslant 

    \fi
  \fi
\fi 

\ifCUPmtlplainloaded \else
  \ifAMStwofonts \else 
    \def\upi{\pi}
    \def\umu{\mu}
    \def\upartial{\partial}
  \fi
\fi

\input psfig
\def\spose#1{\hbox to 0pt{#1\hss}}
\def\lta{\mathrel{\spose{\lower 3pt\hbox{$\sim$}}
    \raise 2.0pt\hbox{$<$}}}
\def\gta{\mathrel{\spose{\lower 3pt\hbox{$\sim$}}
    \raise 2.0pt\hbox{$>$}}}

\title{Asteroids in the Inner Solar System I -- Existence}
\author[S.A. Tabachnik and N.W. Evans]
       {S.A. Tabachnik$^{1,2}$ and N.W. Evans$^1$\\
        $^1$ Theoretical Physics, 1 Keble Rd, Oxford, OX1 3NP\\
        $^2$ Princeton University Observatory, Princeton, NJ
        08544-1001, USA}
\date{}

\begin{document}

\maketitle

\label{firstpage}

\begin{abstract}
Ensembles of in-plane and inclined orbits in the vicinity of the
Lagrange points of the terrestrial planets are integrated for up to
100 million years. The integrations incorporate the gravitational
effects of Sun and the eight planets (Pluto is neglected).  Mercury is
the least likely planet, as it is unable to retain tadpole orbits over
100 million year timescales. Mercurian Trojans probably do not exist,
although there is evidence for long-lived, corotating horseshoe orbits
with small inclinations.  Both Venus and the Earth are much more
promising, as they possess rich families of stable tadpole and
horseshoe orbits. Our survey of Trojans in the orbital plane of Venus
is undertaken for 25 million years. Some $40 \%$ of the survivors are
on tadpole orbits. For the Earth, the integrations are pursued for 50
million years.  The stable zones in the orbital plane are larger for
the Earth than for Venus, but fewer of the survivors ($\sim 20 \%$)
are tadpoles.  Both Venus and the Earth also have regions in which
inclined test particles can endure near the Lagrange points. For
Venus, only test particles close to the orbital plane ($i \lta
16^\circ$) are stable. For the Earth, there are two bands of
stability, one at low inclinations ($i \lta 16^\circ$) and one at
moderate inclinations ($24^\circ \lta i \lta 34^\circ$).  The inclined
test particles that evade close encounters are primarily moving on
tadpole orbits.  Two Martian Trojans (5261 Eureka and 1998 VF31) have
been discovered over the last decade and both have orbits moderately
inclined to the ecliptic ($20.3^\circ$ and $31.3^\circ$
respectively). Our survey of in-plane test particles near the Martian
Lagrange points shows no survivors after 60 million years.  Low
inclination test particles do not persist, as their inclinations are
quickly increased until the effects of a secular resonance with
Jupiter cause de-stabilisation. Numerical integrations of inclined
test particles for timespans of 25 million years show stable zones for
inclinations between $14^\circ$ and $40^\circ$. However, there is a
strong linear resonance with Jupiter which destabilises a narrow band
of inclinations at $\sim 29^\circ$. Both 5261 Eureka and 1998 VF31 lie
deep within the stable zones, which suggests they may be of primordial
origin.

\end{abstract}


%
%

\def\be{\begin{equation}}
\def\ee{\end{equation}}

\section{Introduction}

Lagrange's (1772) triangular solution of the three body problem was
long thought to be just an elegant mathematical curiosity. The three
bodies occupy the vertices of an equilateral triangle. Any two of the
bodies trace out elliptical paths with the same eccentricity about the
third body as a focus (see e.g., Whittaker 1904, Pars 1965).  The
detection of 588 Achilles near Jupiter's Lagrange point in 1906 by
Wolf changed matters. This object librates about the Sun-Jupiter $L_4$
Lagrange point, which is $60^\circ$ ahead of the mean orbital
longitude of Jupiter (e.g., \'Erdi 1997).  About 470 Jovian Trojans
are now known (see
``http://cfa-www.harvard.edu/iau/lists/Trojans.html''), though the
total population exceeding 15 km in diameter may be as high as $\sim
2500$ (Shoemaker, Shoemaker \& Wolfe 1989; French et al. 1989).
Roughly $80 \%$ of the known Trojans are in the $L_4$ swarm. The
remaining $20 \%$ librate about the $L_5$ Lagrange point, which trails
$60^\circ$ behind the mean orbital longitude of Jupiter.  There are
also Trojan configurations amongst the Saturnian moons. The Pioneer 11
and the Voyager 1 and 2 flybys of Saturn discovered five small moons
on tadpole or horseshoe orbits. The small moon Helene librates about
the Saturn-Dione $L_4$ point. The large moon Tethys has two smaller
Trojan moons called Telesto and Calypso, one of which librates about
the Saturn-Tethys $L_4$ and the other about the Saturn-Tethys $L_5$
points. Finally, the two small moons Janus and Epimetheus follow
horseshoe orbits coorbiting with Saturn (see e.g., Smith et al. 1983;
Yoder et al. 1983; Yoder, Synnott \& Salo 1989).  Another example of a
Trojan configuration closer to home is provided by the extensive dust
clouds in the neighbourhood of the $L_5$ point of the Earth-Moon
system claimed by Winiarski (1989).

This paper is concerned with the existence of coorbiting asteroids
near the triangular Lagrange points of the four terrestrial planets.
For non-Jovian Trojans, the disturbing forces due to the other planets
are typically larger than those caused by the primary planet itself.
For this reason, it was formerly considered unlikely that long-lived
Trojans of the terrestrial planets could survive. Over the past
decade, two lines of evidence have suggested that this reasoning is
incorrect. The first is the direct discovery of inclined asteroids
librating about the Sun-Mars $L_5$ Lagrange point. The second is
numerical integrations, which have steadily increased in duration and
sophistication. 

The first non-Jovian Trojan asteroid, 5261 Eureka, was discovered by
Holt \& Levy (1990) near the $L_5$ point of Mars.  Surprisingly, the
orbit of 5261 Eureka is inclined to the plane of the ecliptic by
$20.3^\circ$.  The determination of Eureka's orbital elements and a
preliminary analysis of its orbit were published in Mikkola et
al. (1994).  Numerical integrations were performed for several dozen
Trojan test particles with different initial inclinations for $\sim 4$
Myr by Mikkola \& Innanen (1994), who claimed that long-term stability
of Martian Trojans was possible only in well-defined inclination
windows, namely $15^{\circ} \leq i \leq 30^{\circ}$ and $32^{\circ}
\leq i \leq 44^{\circ}$ with respect to Jupiter's orbit.  The
discovery of a second Mars Trojan, 1998 VF31, soon followed (see e.g.,
{\it Minor Planet Circular 33763}; Tabachnik \& Evans 1999). Perhaps
the most important point to take from the observational discoveries is
that if a comparatively puny body like Mars possesses Trojans, it is
quite likely indeed that the more massive planets also harbour such
satellites.

The main argument for the possible existence of Trojans of the Earth,
Venus and Mercury comes from numerical test particle surveys. Much of
the credit for reviving modern interest in the problem belongs to
Zhang \& Innanen (1988a,b,c). Using the framework of the planar
elliptic restricted four and five-body problem, their integrations
extended from 2000 to $10^5$ years, though their model was not
entirely self-consistent as mutual interactions between the planets
were not taken into account.  Further important results on the
stability of the Trojans of the terrestrial planets were obtained in a
series of papers by Mikkola \& Innanen (1990, 1992, 1994, 1995). The
orbits of the planets from Venus to Jupiter were computed at first
using a Bulirsch-Stoer integration and later with a Wisdom-Holman
(1991) symplectic integrator.  In the case of Mercury, test particles
placed at the Lagrange point exhibited a strongly unstable behaviour
rather rapidly. Conversely, the mean librational motion of all the
Trojan test particles near Venus and the Earth's Lagrange points
appeared extremely stable.  Thus far, the longest available
integrations are the 6 Myr survey of test particles near the
triangular $L_4$ point of Venus and the Earth (Mikkola \& Innanen
1995). The initial inclinations ranged from $0^{\circ}$ to
$40^{\circ}$ with respect to the orbital plane of the primary
planet. The Trojans of Venus and the Earth persisted on stable orbits
for small inclinations ($i \leq 18^{\circ}$ and $i \leq 11^{\circ}$
respectively).

These lines of reasoning suggest that a complete survey of the
Lagrange points of the terrestrial planets is warranted. It is the
purpose of the present paper to map out the zones in which coorbital
asteroids of the terrestrial planets can survive for timescales up to
100 million years.  Section 2 provides a theoretical introduction to
the dynamics of coorbital satellites within the framework of the
elliptic restricted three body problem. Our fully numerical survey is
discussed in Section 3 and takes into account the effects of all the
planets (except Pluto), as well as the most important post-Newtonian
corrections and the quadrupole moment of the Moon. The results of the
survey are presented for each of the terrestrial planets in Sections 4
to 7 (Mercury, Venus, the Earth, and Mars). Finally, a companion paper
in this issue of {\it Monthly Notices} discusses the observable
properties of the asteroids.

\begin{figure}
\centerline{\psfig{figure=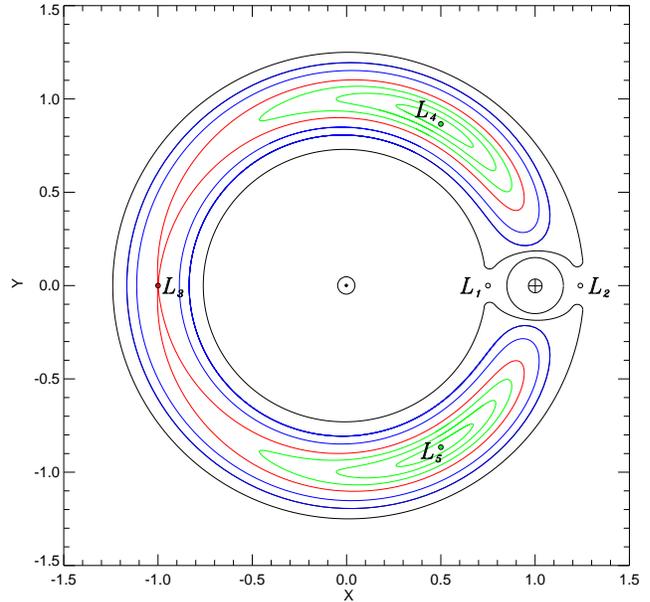,height=\hsize}}
\caption{The positions of the five Lagrange points for the
three body system in a corotating frame.  There are two type of
planar, coorbiting orbits -- the tadpoles and the horseshoes.  Tadpole
orbits are shown in green and librate about either the leading
$\Lfour$ or trailing $\Lfive$ Lagrange points.  Horseshoe orbits are
shown in blue and perform a double libration about both $\Lfour$ and
$\Lfive$. The orbits are based on numerical data for the case of the
Sun, the Earth and an asteroid, but the difference in semimajor axis
of the asteroid and the planet has been magnified by a factor of 40
for clarity. The dividing separatrix or orbit of infinite period is
shown in red. The boundaries of the coorbital region are shown as full
lines.  They are schematic and correspond to bounding positions
limited by $L_1$ and $L_2$, which are themselves arbitrarily
located. The satellite r\'egime occupies a sphere around the planet.}
\label{fig:orbitplot}
\end{figure}
\begin{figure}
\centerline{\psfig{figure=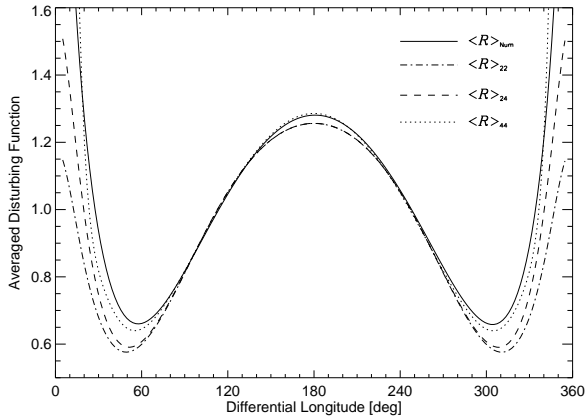,height=0.7\hsize}}
\caption{The averaged disturbing function~(\ref{eq:totalavpot})
plotted as a function of differential longitude $\phi$. The figures
shows approximations carried out to second order in inclination and
eccentricities $\langle R \rangle_{22}$, second order in inclination
and fourth order in eccentricity $\langle R \rangle_{24}$ and fourth
order in both quantities $\langle R \rangle_{44}$. They are compared
to the numerically evaluated averaged disturbing function $\langle R
\rangle_{\rm Num}$.}
\label{fig:potentialplot}
\end{figure}
\begin{table}
\begin{center}
\begin{tabular}{|c|c|c|c|} \hline
Planet & $\Lfour$ & $\Lfive$ & $\phi_\star$ \\ \hline
Mercury & $59.822^\circ$ & $300.178^\circ$ & $25.736^\circ$ \\ \hline
Venus & $60.000^\circ$ & $300.000^\circ$ & $23.908^\circ$ \\ \hline
Earth & $59.999^\circ$ & $300.001^\circ$ & $23.917^\circ$  \\ \hline
Mars & $59.964^\circ$ & $300.036^\circ$ & $24.266^\circ$  \\ \hline
\end{tabular}
\end{center}
\caption{The angular positions of the Lagrange points as inferred from
(\ref{eq:totalavpot}) for each of the terrestrial planets. The
asteroidal orbit is assumed to have vanishing inclination and argument
of pericentre.  Note that the Lagrange points are displaced only
slightly from their classical values. Also recorded is the angular
position of the separatrix, which physically represents the minimum
differential longitude that can be reached by the tadpole orbits.}
\label{table:separatrixposition}
\end{table}

\section{Theoretical Framework}

\noindent
Consider the restricted three-body problem, defined by a system of two
massive bodies on Keplerian orbits attracting a massless test particle
that does not perturb them in return. The five Lagrange points are the
stationary points of the effective potential (e.g., Danby 1988). The
collinear Lagrange points $L_1$, $L_2$ and $L_3$ are unstable, whereas
the triangular Lagrange points $\Lfour$ and $\Lfive$ form equilateral
triangles with the two massive bodies, as illustrated in
Fig.~\ref{fig:orbitplot}.  Motion around $\Lfour$ and $\Lfive$ can be
stable.  The planar orbits that corotate with a planet are classified
as either tadpoles or horseshoes. Tadpole orbits perform a simple
libration about either the Lagrange point $\Lfour$ or $\Lfive$,
whereas the horseshoes perform a double libration about both $\Lfour$
and $\Lfive$. The tadpole and horseshoe orbits are divided by a
separatrix, which corresponds to an orbit of infinite period that
passes through $\Lthree$.  On average, both tadpoles and horseshoes
move about the Sun at the same rate as the parent planet.

It is helpful to develop an approximate treatment of coorbital motion.
Within the framework of the elliptic restricted problem of three
bodies (Sun, planet and asteroid), the Hamiltonian of the asteroid in a
non-rotating heliocentric coordinate system reads:
\begin{equation}
H =  -\frac{k^2}{2a} - \mplanet k^2 R,
\end{equation}
where $\mplanet$ is the mass of the planet in Solar masses and $k$ is the
Gaussian gravitational constant. The disturbing function $R$ is
defined by (e.g., Danby 1988)
\begin{eqnarray}
R &=& \frac{1}{(r^2+{\rplanet}^2-2r\rplanet\cos S)^{1/2}} - \frac{r \cos
S}{{\rplanet}^2}.
\end{eqnarray} 
Here, the radius vector $r$ and $\rplanet$ refer to the asteroid and the
planet respectively, $S$ being the elongation of the asteroid from the
parent planet. Following Brouwer \& Clemence (1961), $\cos S$ can be
expressed in function of the mutual inclination of the two orbits $i$,
the true anomalies $\nu$ and $\nuplanet$, the argument of perihelion
$\omega$ and the longitude of the ascending node $\Omega$ of the
asteroid:
\begin{eqnarray}
\cos S & =&  \cos ^2 \halfi\cos (\nu - \nuplanet + \omega + \Omega) 
\nonumber \\
& + & \sin ^2 \halfi \cos (\nu + \nuplanet + \omega - \Omega).
\end{eqnarray}
The disturbing function is then expressed in terms of the mean synodic
longitudes $\lambdaplanet=\Mplanet$ and
$\lambda=M+\omega+\Omega$. Here, $\Mplanet$ and $M$ are the mean
anomalies of the planet and asteroid. In other words, we are using a
coordinate system for which $(x,y)$ lies in the orbital plane of Mars
and the $x$-axis points towards Mars' perihelion.  The disturbing
function is expanded to second order in the eccentricities and to
fourth order in the inclination. Finally, we make a change of
variables $\phi=\lambdaplanet -\lambda$ and average $R$ over the mean
anomaly of the planet $\Mplanet$:
\begin{equation}
\langle R \rangle = {1\over 2 \pi} \int_0^{2\pi} R\, d \Mplanet =
U_0 + U_2 + O(e^3, \eplanet^3, i^5).
\label{eq:totalavpot}
\end{equation}
The zeroth order term in the averaged disturbing function is
\begin{equation}
U_0 = {1\over (a^2 + \aplanet^2 -2a\aplanet \cos^2 \halfi \cos \phi)^{1/2}} 
-{a\over \aplanet^2}\cos^2 \halfi\cos \phi. 
\label{eq:zeroth}
\end{equation}
The term that is second order in the eccentricities and inclinations
has three parts
\begin{eqnarray}
U_2 = U_{2,0} + {U_{2,3}\over (a^2 + \aplanet^2 -2a\aplanet \cos^2
 \halfi \cos \phi)^{3/2}} \nonumber \\
+ {U_{2,5}\over 
(a^2 + \aplanet^2 -2a\aplanet \cos^2 \halfi \cos \phi)^{5/2}},
\label{eq:potential}
\end{eqnarray} 
where $U_{2,0}, U_{2,3}$ and $U_{2,5}$ are given in Appendix A. Still
lengthier expressions accurate to the fourth order in the
eccentricities and inclinations are given in Tabachnik (1999). It is
interesting to note that the second order expressions depend only on
the longitude of perihelion, while the fourth order contains terms
depending on both the longitude of perihelion and the longitude of the
ascending node.  Some of these averaged disturbing functions are shown
in Fig.~\ref{fig:potentialplot} and compared to the numerically
computed and averaged disturbing function $\langle R \rangle_{\rm Num}$.

For small and moderate inclinations, $\langle R \rangle$ has a
double-welled structure. The libration about the Lagrange points can be
viewed as the trapping of the test particle in the well, with the
tadpole orbits located in the deeper part of the well and the
horseshoe orbits in the shallower part. The two orbital families are
divided by a separatrix.  The slight eccentricities and inclinations
of the planets cause small deviations of the Lagrange points from
their classical values (c.f. Namouni \& Murray 1999). These deviations
can be calculated by finding the local minima of the secular
potential~(\ref{eq:totalavpot}). The results for each terrestrial
planet are recorded in Table~\ref{table:separatrixposition}.

The position of the separatrix is found by locating the local maximum
of the secular potential~(\ref{eq:totalavpot}). The separatrix
terminates at a differential longitude from the planet
$\phi_\star$. This angle is the closest any tadpole orbit can come to
the planet. In the restricted circular problem, this angle is
$\phi_\star = 2\asin (\ffrac{1}{\sqrt{2}} -\ffrac{1}{2}) =
23.906^\circ$ (see e.g., Brown \& Shook 1933). The deviation from this
classical value caused by the eccentricity and inclination of the
planets is also recorded in Table~\ref{table:separatrixposition}.

The secular potential~(\ref{eq:totalavpot}) can also be used to work
out the approximate period $P$ of small angle librations about the
Lagrange points. For simplicity, we take only the zeroth order
term~(\ref{eq:zeroth}) with $a = \aplanet$ and obtain the minimum of
the secular potential at
\begin{equation}
\phi = \acos \Bigl( \frac{1}{2 \cos^2 \halfi} \Bigr).
\end{equation}
The period of small librations about this minimum is
\begin{equation}
P = \frac{2}{3} {\Pplanet \over \sqrt{\mplanet (4 \cos^4(\halfi) -1)}}
\end{equation}
where $\Pplanet$ is the orbital period of the planet. The period
of libration increases with increasing inclination.

Expansion of the disturbing function is a good guide to the dynamics
provided the inclinations and eccentricities are low or moderate. At
high inclinations and eccentricities, the classification of the
corotating orbits has only recently been undertaken by Namouni (1999).
\begin{table*}
\begin{center}
\begin{tabular}{|c|c|c|c|c|c|c|c|c|c|} \hline
& Hyp & Mer & Ven  & Ear & Mar & Jup & Tad & Hor & Total \\ \hline
Mercury & 6 & 572 & 157 & 1 & 0 & 3 & 0 & 53 & 792\\ \hline
Venus & 0 & 0 & 385 & 0 & 0 & 0 & 168 & 239 & 792\\ \hline
Earth & 1 & 0 & 2 & 280 & 0 & 0 & 95 & 414 & 792\\ \hline
Mars & 1 & 0 & 0 & 27 & 763 & 1 & 0 & 0 & 792\\ \hline
\end{tabular}
\end{center}
\caption{Statistics for test particles in the plane of each parent
planet. The second column lists the number of test particles removed
from the integration because the orbit becomes hyperbolic. The next
columns give the number of close encounters with each named
planet. The number of survivors on tadpole and horseshoe orbits 
respectively are also recorded.}
\label{table:statisticsinplane}
\end{table*}
\begin{table*}
\begin{center}
\begin{tabular}{|c|c|c|c|c|c|c|c|c|c|c|c|c|} \hline
        & Hyp & Mer  & Ven & Ear & Mar & Jup & Sat & Ura & Nep & Tad & Hor
& Tot \\ \hline
Mercury & 320 & 558  & 132 & 40  & 1   & 33  & 6   & 1   & 0   & ??(0)&
??(13) & 1104\\ \hline
Venus   & 32  & 19   & 695 & 209 & 1   & 6   & 5   & 0   & 0   & 129    &
8 & 1104\\ \hline
Earth   & 44  & 0    & 171 & 669 & 9   & 7   & 3   & 0   & 1   & 182  & 18
& 1104\\ \hline
Mars    & 99  & 0    & 75  & 311 & 445 & 12  & 4   & 0   & 1   & 157  & 0
& 1104\\ \hline
\end{tabular} 
\end{center}
\caption{As table~\ref{table:statisticsinplane}, but for test
particles inclined with respect to the plane of the parent planet.
There are 1104 test particles initially. The numbers terminated
because their orbits become hyperbolic or because they enter the
sphere of influence of any of the planet are reported. The number of
survivors on tadpole and horseshoe orbits respectively are also
recorded. There is some uncertainty as to the number of surviving
highly inclined test particles around Mercury, as explained in the
text.}
\label{table:statisticsinclined}
\end{table*}

\begin{figure*}
\centerline{\psfig{figure=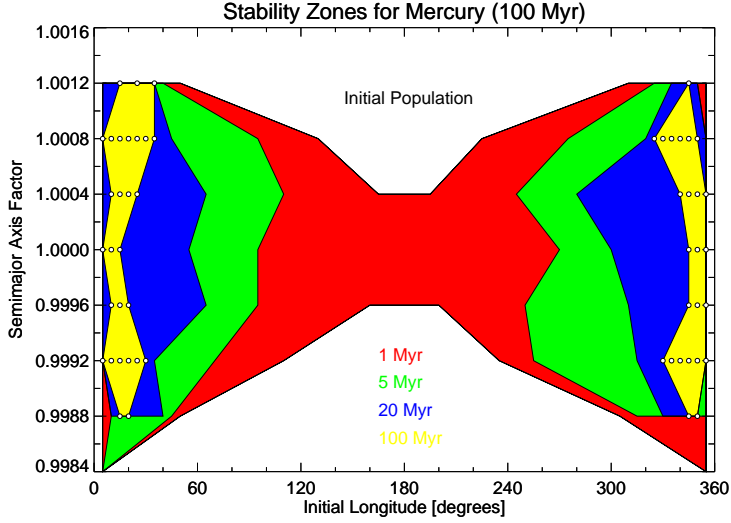,height=0.4\hsize}}
\caption{This shows the gradual erosion of test particles in the
vicinity of the Lagrange points of Mercury. The initial population of
test particles covers every $5^\circ$ in longitude and every $0.0004$
in semimajor axis factor. Red, green, blue and yellow mark the
positions of the survivors after 1, 5, 20 and 100 million years.  }
\label{fig:merccolorplot}
\end{figure*}
\begin{figure*}
\centerline{\psfig{figure=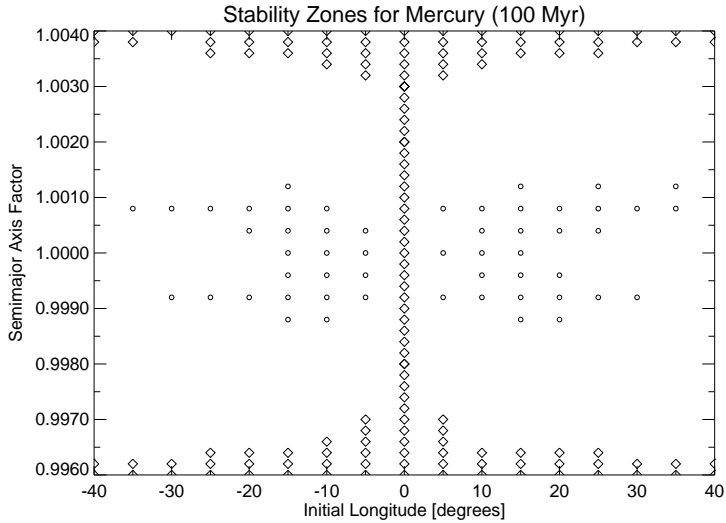,height=0.4\hsize}}
\caption{This shows the surviving test particles in the plane
of initial longitude and semimajor axis factor marked as circles. The
surviving orbits are all horseshoes.  The length of integration is 100
million years.  Unstable test particles in the elliptic restricted
three body problem (comprising the Sun, Mercury and test particle) are
shown as diamonds.  Everything within the inner boundary of diamonds
is stable at the level of the elliptic restricted three body problem,
and so this indicates the damaging effect of perturbations from the
rest of the Solar System.}
\label{fig:mercbwplot}
\end{figure*}
\begin{figure*}
\centerline{\psfig{figure=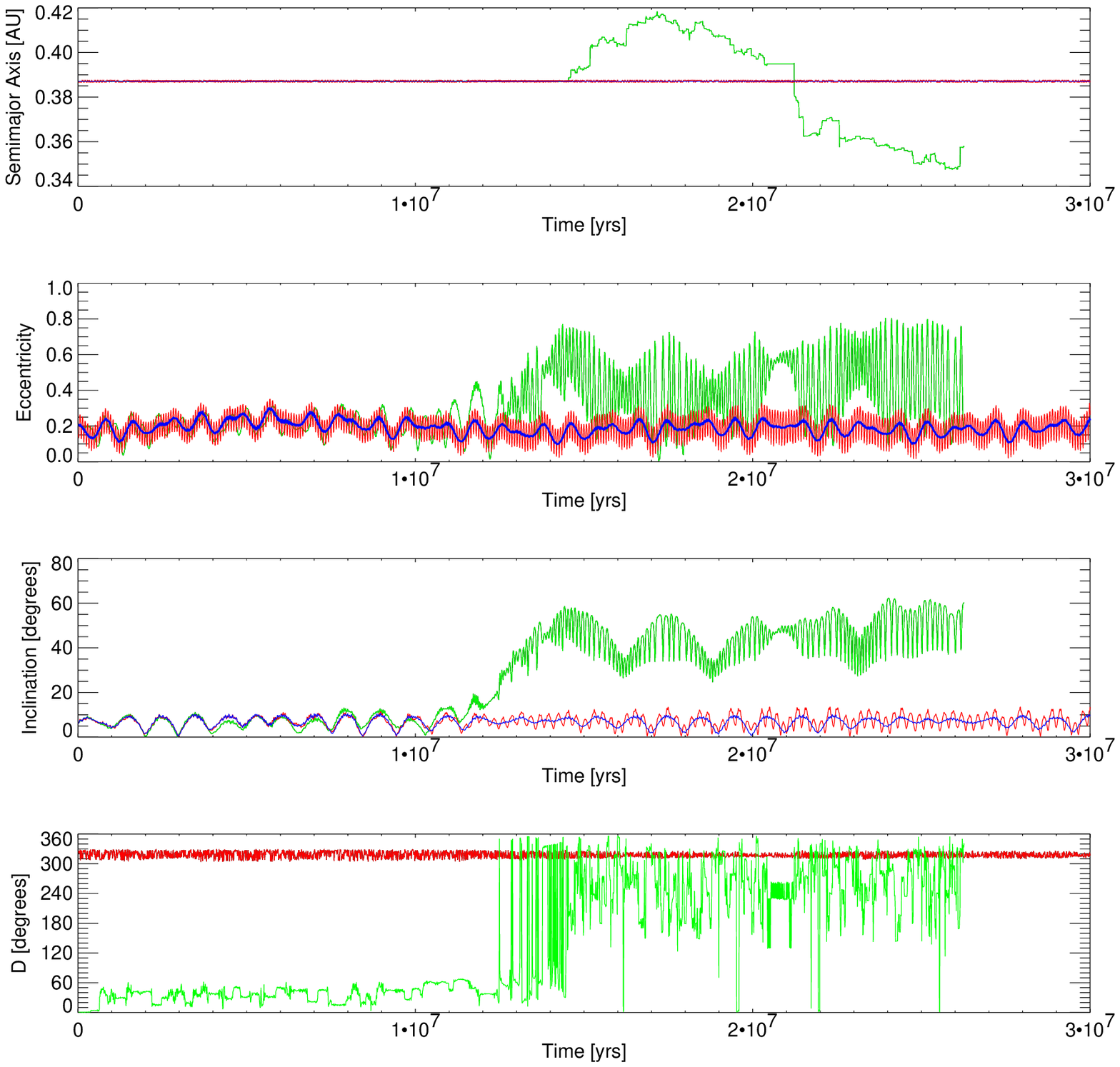,height=\hsize}}
\caption{The first three panels of the figure show the evolution of
the semimajor axis, eccentricity and inclination for Mercury (blue), a
stable test particle (red) and an unstable test particle (green).  The
final panel shows the amplitude of libration $D$ about the Lagrange
point. Note that the stable test particle follows a horseshoe orbit,
as evidenced by $D \sim 320^\circ$. The unstable test particle is
initially on a tadpole orbit, then passes through a brief horseshoe
phase before entering the sphere of influence of Mercury.}
\label{fig:mercorbits}
\end{figure*}
\section{Numerical Method}

\noindent
For each of the terrestrial planets, we carry out numerical surveys of
in-plane and inclined corotating orbits.  Section 3.1 introduces the
integration method, while Section 3.2 discusses the numerical
procedure.

\subsection{Mixed Variable Symplectic Integrators}

\noindent
Our model includes all the planets (except Pluto whose contribution is
negligible in the inner Solar System). The asteroids are represented
as test particles with infinitesimal mass. They are perturbed by the
planets but they do not perturb them in return.
 
The orbits of the planets are integrated using a mixed variable
symplectic integrator scheme (Wisdom \& Holman 1991; Kinoshita,
Yoshida \& Nakai 1991) with individual time steps (Saha \& Tremaine
1994) which takes into account post-Newtonian corrections and the
quadrupole moment of the Sun's attraction on the barycentre of the
Earth-Moon system (Quinn, Tremaine \& Duncan 1991). The two latter
contributions must be included for calculations in the inner Solar
System, whereas the cumulative effect of the asteroids, satellites,
galactic tidal acceleration, passing stars, solar mass loss and
oblateness is believed to be smaller than $\sim 10^{-10}$ and is
neglected (Quinn et al. 1991). Mixed variable symplectic integrators
exploit the fact that the Hamiltonian written in Jacobi coordinates
(Plummer 1960; Wisdom \& Holman 1991) is dominated by a nearly
Keplerian term. The mixed variable symplectic integrators are
so-called because they evaluate the planetary disturbing forces in
Cartesian coordinates while using the elements to advance the
orbits. Fast algorithms, like Gauss' $f$ and $g$ functions, exist to
perform the latter task (e.g., Danby 1988; Wisdom \& Holman 1991).

The orbit of any of the test particles is derived from the Hamiltonian
\begin{equation} 
H_{\rm{tp}} = H_{\rm{kep}} + H_{\rm{int}}.
\end{equation}
Here, the test particle is given the first Jacobi index and the
planets carry the higher Jacobi indices. As Wisdom \& Holman (1991)
point out, this is an advantageous choice as it gives the simplest
interaction Hamiltonian.  Denoting the Jacobi position of the test
particle as ${\bf x}_1$ and its velocity as ${\bf v}_1$, then
\begin{equation}
\label{eqn:htp}
H_{\rm{kep}} =  {v_1^2\over 2} - {k^2\over r_1}, \qquad
H_{\rm{int}} = k^2 \sum_{i=2}^{N+1}  m_i\left( \frac{{\bf x}_1 \cdot
{\bf x}_i}{r_i^3} - \frac{1}{r_{1i}}\right),
\end{equation}
where $k$ is the Gaussian gravitational constant and $N$ is the number
of planets included in the model ($N=8$ for our calculations). We have
used the notation $r_{1i} = |{\bf x}_1 - {\bf x}_i |$ as the distance
from the test particle to the $i$th body.  As mentioned by Wisdom and
Holman (1991), the intuitive interpretation of $H_{\rm int}$ is to
note that the attraction of the Sun on the test particle equals the
difference between the direct acceleration of the massive planets on
the test particle and the gravitational pull of the planets on the
Sun.

The most important general relativistic effects can be included by
modifying the test particle Hamiltonian to
\begin{equation} 
H_{\rm{tp}} = H_{\rm{kep}} + H_{\rm{int}} + H_{\rm PN}.
\end{equation}
A clever device for incorporating the most important post-Newtonian 
effects into mixed variable symplectic integrators is given by Saha
\& Tremaine (1994). The post-Newtonian Hamiltonian is recast as
\begin{equation}
\label{eqn:hamtp}
H_{\rm{PN}} = \frac{1}{c^2} \left( \frac{3}{2} H_{\rm{kep}}^2 -
\frac{k^4}{r_1^2} - \frac{v_1^4}{2} \right).
\end{equation}
The last expression contains three terms which are each integrable
individually. The Keplerian part of eq.~(\ref{eqn:hamtp}) can be
incorporated into the usual Keplerian orbital advance using
\begin{equation}
\label{eqn:exp}
\exp \left( \tau \lbrace \;\;, H_{\rm{kep}} + \frac{3}{2c^2}
H_{\rm{kep}}^2 \rbrace \right). 
\end{equation}
The Keplerian Hamiltonian is conserved with time and equals
$-\frac{1}{2} k^2/a_{\rm{tp}}$, $a_{\rm{tp}}$ being the semi-major
axis of the test particle. After some straightforward algebra,
eq.~(\ref{eqn:exp}) becomes
\begin{equation}
\exp \left( \tau ' \lbrace \;\;, H_{\rm{kep}}\rbrace \right),
\;\;\; \tau ' = \left( 1 - \frac{3k^2}{2c^2 a_{\rm{tp}}} \right)
\tau.
\end{equation}
The second and third terms within the bracket of eq.~(\ref{eqn:hamtp})
imply modifications of the test particle position and velocity before
and after advancing the Keplerian orbit:
\begin{equation}
\frac{d {\bf v}_1}{dt} = - \frac{2k^4}{c^2} \frac{{\bf r}_1}{r_1^4},\qquad
\frac{d {\bf x}_1}{dt} = - \left( \frac{2}{c^2} v_1^2 \right) {\bf v}_1
\end{equation}
Further details of this neat formulation are given in Saha \& Tremaine
(1994).

\subsection{The Numerical Procedure}

\noindent
For each of the terrestrial planets, we carry out two surveys. The
first is restricted to test particles in the orbital plane of the
planet.  The test particles are given the same eccentricity $e$,
inclination $i$, longitude of the ascending node $\Omega$ and mean
anomaly $M$ as the planet. The argument of pericentre $\omega$ is
varied from $0^\circ$ to $360^\circ$ in steps of $5^\circ$.  The
initial semimajor axis is equal to the semimajor axis of the planet
multiplied by a semimajor axis factor (c.f., Innanen \& Mikkola's
(1989) investigation of Saturnian Trojans).  The second survey is
restricted to test particles with the same semimajor axis as the
planet. The initial inclinations of the test particles (with respect
to the plane of the planet's orbit) are spaced every $2^\circ$ and the
initial arguments of pericentre are spaced every $15^\circ$. The
eccentricities of the asteroids are inherited from the parent planet.

The initial conditions come from the JPL Planetary and Lunar
Ephemerides, DE405 which is available at ``http://ssd.jpl.nasa.gov/''
(Chamberlain et al. 1997). The starting epoch of the integration is
JED2440400.5 (28 June 1969). The standard units used for the
integration are the astronomical unit, the day and the Gaussian
gravitational constant $k^2 = GM_\odot$. The Earth to Moon mass ratio
is $M_{\oplus}/M_{\rm L} = 81.3$. For most of the computations
described below, the timestep for Mercury is $14.27$ days. The
timesteps of the planets are in the ratio $1:2:2:4:8:8:64:64$ for
Mercury moving outwards, so that Neptune has a timestep of $2.5$
years. The test particles all have the same timestep as Mercury.
These values were chosen after some experimentation to ensure the
relative energy error has a peak amplitude of $\approx 10^{-6}$ over
the tens of million year integration timespans (c.f., Holman \& Wisdom
1993; Saha \& Tremaine 1994).  Individual planetary stepsizes do
introduce additional stepsize resonances and it is important to check
that the resonances do not degrade the accuracy of the numerical
results. Referring to Figure 3 of Wisdom \& Holman (1992), the
stepsize resonances begin to overlap when the ratio of semimajor axes
is $\gta 0.8$ and stepsize is $\gta 0.2$ times the orbital
period. Bearing this in mind, it seems that the stepsize resonances
are not a serious concern even for the test particles near Mercury.
Nonetheless, Mercury does provide severe challenges for long-term
integrations, and there is some evidence that round-off error may be
affecting our results for the highly inclined test particles around
Mercury.

As the test particles' orbits are integrated, they are examined at
each time step. If their trajectories become parabolic or hyperbolic
orbits, they are removed from the survey. In addition, test particles
which experience close encounters with a massive planet or the Sun are
also terminated. The sphere of influence is defined as the surface
around a planet at which the perturbation of the planet on the
two-body heliocentric orbit is equal to that of the Sun on the
two-body planetocentric orbit (e.g., Roy 1988). If the planet's mass
is much less than that of the Sun, this surface is roughly spherical
with radius
\begin{equation}
r_{\rm s} = \aplanet m_{\rm p}^{2/5},
\end{equation}
where $a_{\rm p}$ is the semi-major axis of the planet and $m_{\rm p}$
is the planet's mass in solar mass units. Since the algorithm we use
does not allow the variable step size necessary for the treatment of
close encounters, the exact size of the sphere of influence is not of
great importance. Furthermore, test particles which enter the sphere
of influence are typically ejected from the solar system in another
1-10 Myr (e.g., Holman 1997). In the case of the Sun, a close
encounter is defined to be passage within 10 solar radii ($\simeq
0.005$ AU). This general procedure is inherited from a number of
recent studies on the stability of test particles in the Solar System
(see e.g., Gladman \& Duncan 1990; Holman \& Wisdom 1993; Holman 1997;
Evans \& Tabachnik 1999)

We discuss the detailed results for each of the terrestrial planets in
turn. A broad overview of the results is given in three tables. The
fates of the test particles in the in-plane and inclined surveys are
summarised in Tables~\ref{table:statisticsinplane}
and~\ref{table:statisticsinclined} respectively. These provide the
number of survivors on tadpole and horseshoe orbits, as well as the
numbers suffering close encounters with each
planet. Table~\ref{table:tableofeandi} provides the average
eccentric-ties and inclinations of the survivors at the end of the
simulations.

\begin{figure*}
\centerline{\psfig{figure=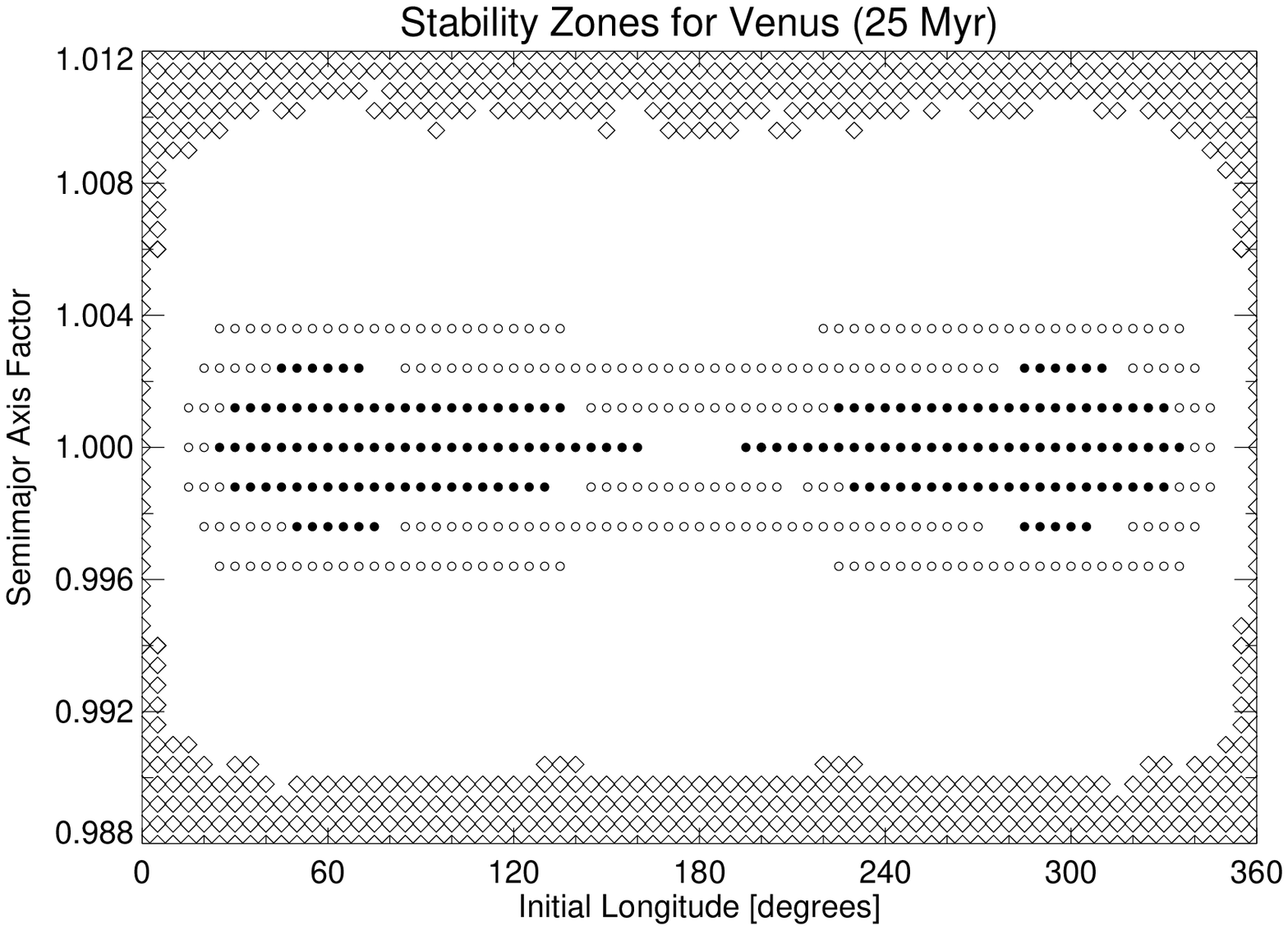,height=0.4\hsize}}
\caption{This shows the surviving test particles near Venus. The
survivors are plotted as circles in the plane of initial longitude and
semimajor axis factor. Filled circles are tadpole orbits, open circles
are horseshoe orbits. Unstable test particles in the restricted three
body problem comprising the Sun, Venus and asteroid are shown as
diamonds. The length of integration is 25 million years.}
\label{fig:venusplaneplot}
\end{figure*}
\begin{figure*}
\centerline{\psfig{figure=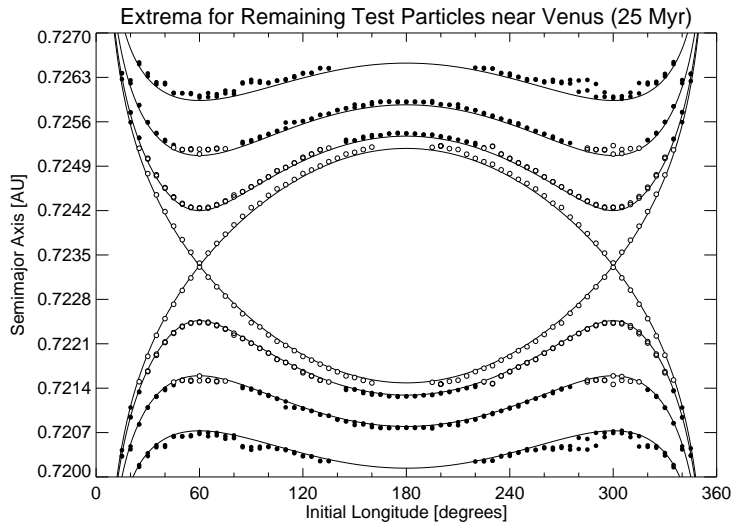,height=0.4\hsize}}
\caption{The extrema of the semimajor axes of the surviving test
particles near Venus are plotted against their initial longitude.  The
extrema of tadpole orbits are shown as filled circles, the extrema of
horseshoes as open circles.  The extrema of survivors starting with
the same semimajor axis factor fall on the same curve
(eq.~\ref{eq:extrema}), which is drawn as a full line.  This is really
a consequence of the fact that the motion is derivable from an
integrable Hamiltonian to an excellent approximation.}
\label{fig:venusextremaplot}
\end{figure*}
\begin{figure*}
\centerline{\psfig{figure=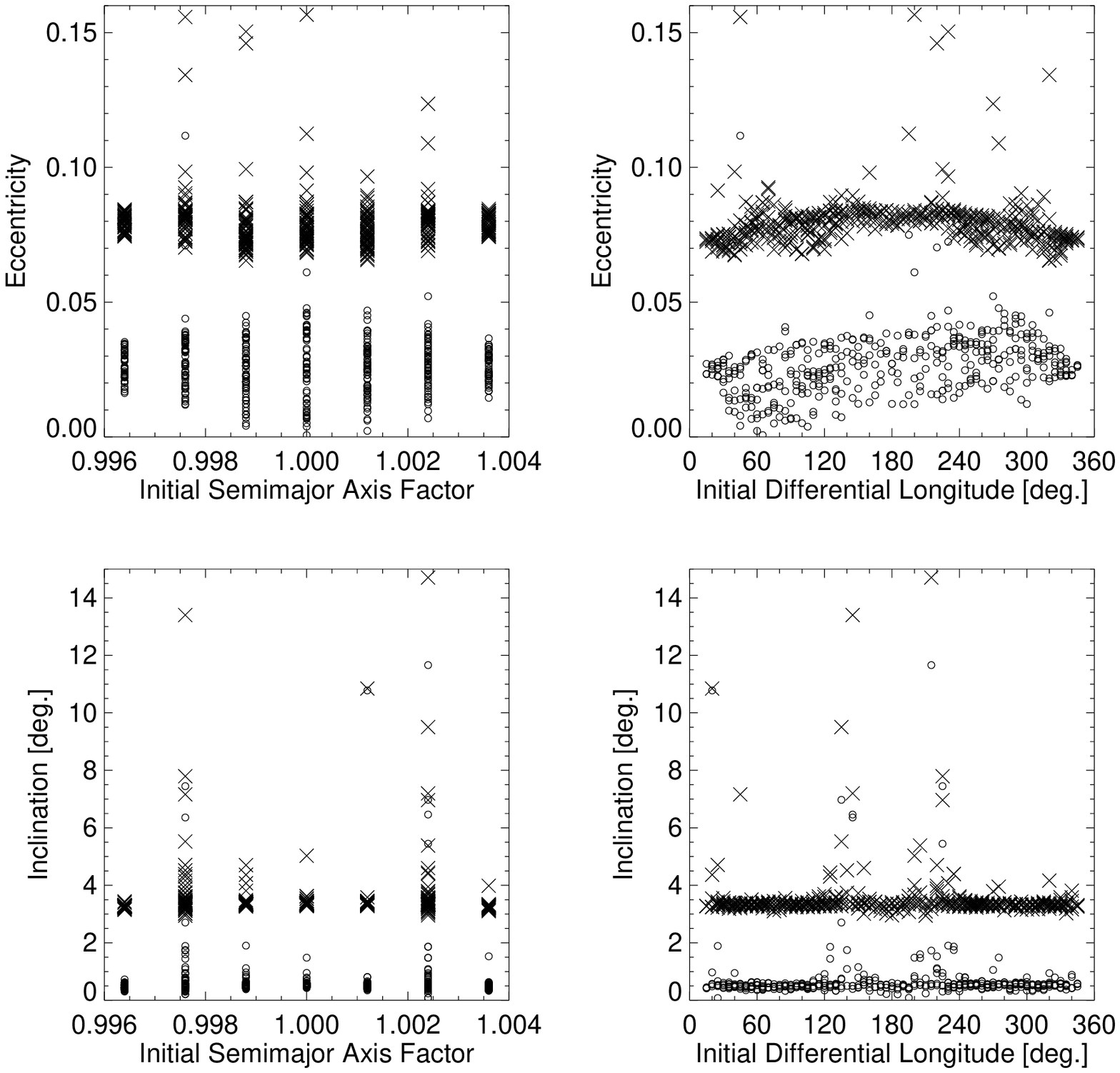,height=0.8\hsize}}
\caption{This shows the distributions of eccentricities and
inclinations of the surviving test particles of
Fig.~\ref{fig:venusplaneplot} against their initial semimajor axis and
longitude from Venus. In all the panels, the maximum values attained
during the course of the orbit integrations are marked by crosses, the
instantaneous values after 25 million years are marked by circles.
The mean maximum eccentricity is $0.079$, while the mean maximum
inclination is $3.518^\circ$. At the end of the simulation, the
average eccentricity and inclination of the sample is $0.027$ and
$0.706^\circ$ respectively. The survivors are members of a low
eccentricity and low inclination family.}
\label{fig:venuspanelsplot}
\end{figure*}

\section{Mercurian Surveys}

For the Mercurian in-plane survey, the semimajor axis factor is chosen
between $0.998$ and $1.002$ in steps of $0.0004$. If the semimajor
axis factor is exactly unity and the argument of pericentre is
displaced by $60^\circ$ or $300^\circ$, then the test particle is at
the classical Lagrange point and can remain there on a stable orbit if
the perturbations from the rest of the Solar System are
neglected. Fig.~\ref{fig:merccolorplot} shows the time evolution of
this array of test particles. The regions occupied by the test
particles remaining after 1, 5, 20 and 100 million years are shown in
red, green, blue and yellow respectively. After 100 million years,
only 53 out of the original 792 test particles remain.  The locations
of the survivors are shown in close-up in
Fig.~\ref{fig:mercbwplot}. The most striking point to notice is that
the stable zones do not include the classical Lagrange points
themselves. In fact, all the survivors follow horseshoe orbits and
there are no surviving tadpole orbits. There are no long-lived
Mercurian Trojans.  There are two possible reasons for this. First,
Mercury is the least massive of the terrestrial planets and therefore
the potential wells in which any long-lived Trojans inhabit are less
deep than for the other terrestrial planets. Second, Mercury is the
most eccentric of the terrestrial planets and this also encourages the
erosion of the test particles. During the course of the 100 million
years simulation, Mercury's eccentricity fluctuates between $\sim 0.1$
and $\sim 0.3$.

The top three panels of Fig. \ref{fig:mercorbits} show the evolution
of the semimajor axis, inclination and eccentricity for Mercury,
together with a stable and an unstable test particle.  The
distributions of the orbital elements of the survivors suggest that
they belong to a low inclination, low eccentricity family of horseshoe
orbits.  At the end of the simulation, the average eccentricity of the
survivors is $0.186$ and their average inclination is $6.968^\circ$.
The lowest panel shows the evolution of the amplitude of libration $D$
(in degrees) about the Lagrange point for the two test particles.  The
stable particle starts off at a semimajor axis factor of 1.001 and an
initial longitude of $15^\circ$. The unstable particle starts off at
exactly the $L_5$ Lagrange point. Notice that the eccentricity and
inclination variations of the stable particle closely follow those of
Mercury. The unstable test particle maintains its tadpole character
only for some $\sim 1.2 \times 10^7$ years, before passing through a
brief horseshoe phase for $\sim 0.5 \times 10^7$ years. After this,
its semimajor axis increases to $\sim 0.42$ AU and finally decreases
to $\sim 0.35$ AU before entering the sphere of influence of
Mercury. The stable test particle follows a horseshoe orbit. This is
obvious on examining the lowest panel, which shows that its angular
amplitude of libration about the Lagrange point is $\sim 320^\circ$.

For the Mercurian inclined survey, the orbits of 1104 test particles
around Mercury are followed for 100 million years.  Only thirteen test
particles survived until the end of the 100 million year
integration. Seven of these were low inclination ($i < 6^\circ$) test
particles started off at arguments of pericentre ($\omega = 15^\circ$
or $345^\circ$) very close to that of Mercury. The remaining six
survivors were all high inclination objects. On repeating these
calculations on similar machines with identical roundoff, but an
updated ephemerides, all six of these highly inclined orbits were
terminated before the end of the 100 million years (mostly because they
entered the sphere of influence of Mercury), although the low
inclination results were reproduced successfully.  Clearly, it is not
possible to draw definitive conclusions about the longevity of highly
inclined asteroids near Mercury, although it seems likely that any
stable zones must be small and depend sensitively on initial
conditions.
\begin{table}
\begin{center}
\begin{tabular}{|c|c|c|c|c|} \hline\hline
Planet & $\langle e \rangle$ & $\langle i \rangle$ & 
$\langle \emax \rangle $ & $\langle \imax \rangle$ \\ \hline\hline
\multicolumn{5}{c} {In-plane survey} \\\hline\hline
Mercury & $0.186$ & $6.968^\circ$ & $0.352$ & $13.307^\circ$ \\ \hline
Venus & $0.027$ & $0.706^\circ$ & $0.079$ & $3.518^\circ$ \\ \hline
Earth & $0.038$ & $1.349^\circ$ & $0.086$ & $3.213^\circ$ \\ \hline
Mars & - & - & - & - \\ \hline\hline
\multicolumn{5}{c} {Inclined survey} \\\hline\hline
Mercury & $0.400$ & $34.069^\circ$ & $0.626$ & $43.945^\circ$ \\ \hline
Venus & $0.041$ & $6.941^\circ$ & $0.122$ & $10.084^\circ$ \\ \hline
Earth & $0.064$ & $15.778^\circ$ & $0.129$ & $17.936^\circ$ \\ \hline
Mars & $0.103$ & $25.858^\circ$ & $0.173$ & $24.267^\circ$ \\ \hline
\end{tabular}
\end{center}
\caption{This lists the average eccentricity $\langle e \rangle$ and 
inclination $\langle i \rangle$ of the surviving test particles at the
end of the simulation. Also recorded is the mean of the maxima of the
eccentricities $\langle \emax \rangle$ and inclination $\langle \imax
\rangle$ during the course of the simulation.
}
\label{table:tableofeandi}
\end{table}

\begin{figure*}
\centerline{\psfig{figure=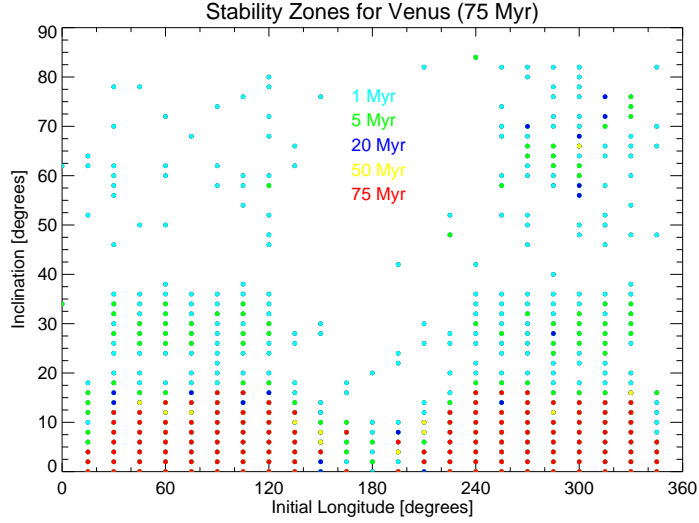,height=0.4\hsize}}
\caption{This shows the erosion of an ensemble of inclined test
particles positioned at the same semimajor axis as the Lagrange point
of Venus but varying in longitude. The test particles surviving after
5, 20, 50 and 75 million years are shown in green, blue, yellow and
red respectively. Unlike the case of Mercury, there are zones of
long-lived stability which hug the the line of vanishing inclination
(to the plane of Venus' orbit).}
\label{fig:venusinclineplotcol}
\end{figure*}
\begin{figure*}
\centerline{\psfig{figure=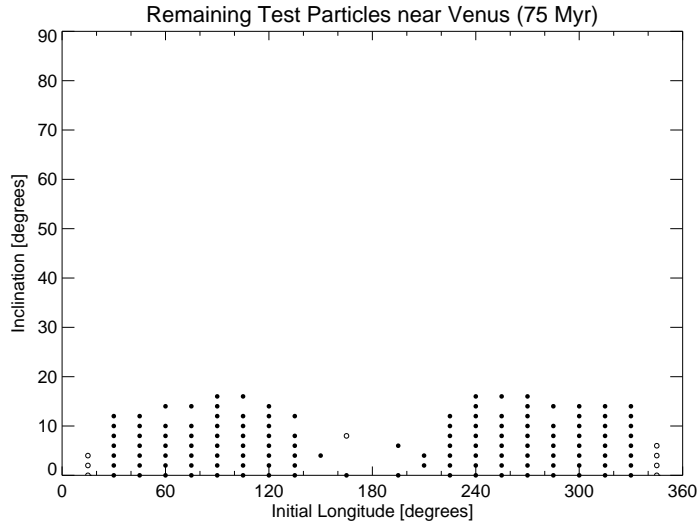,height=0.4\hsize}}
\caption{This shows a close-up of the stability zones of the inclined
Venusian survey. Only test particles surviving the full integration
time of 75 million years are plotted. The filled circles are tadpole
orbits, while the open circles are horseshoe orbits. Moderate or
highly inclined ($i \gta 16^\circ$) Venusian asteroids are unstable,
but this figure provides evidence in favour of a stable population of
low inclination Venusian asteroids.}
\label{fig:venusinclineplotbw}
\end{figure*}
\begin{figure*}
\centerline{\psfig{figure=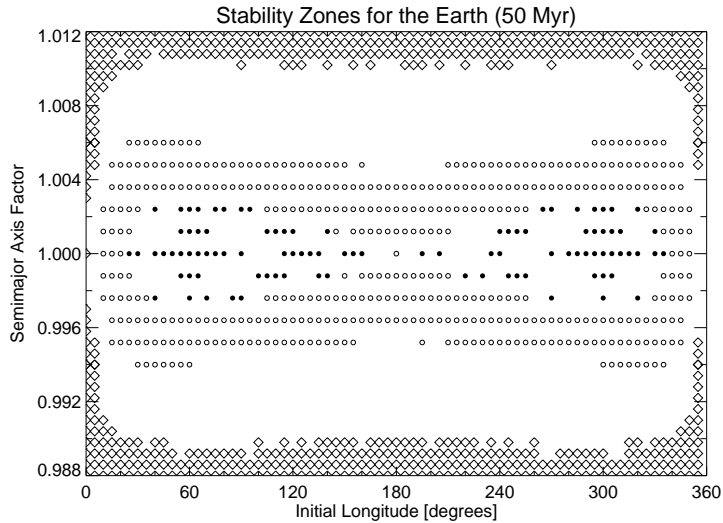,height=0.4\hsize}}
\caption{This shows the surviving in-plane test particles near the
Earth. Filled circles are tadpole orbits, open circles are horseshoe
orbits. Unstable test particles in the restricted three body problem
are shown as diamonds. The length of integration is 50 million years.}
\label{fig:earthplaneplot}
\end{figure*}
\begin{figure*}
\centerline{\psfig{figure=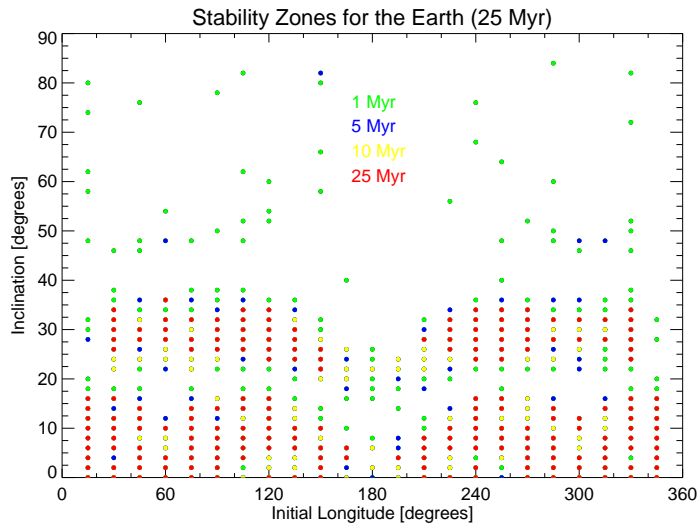,height=0.4\hsize}}
\caption{This shows the erosion of an ensemble of inclined test
particles positioned at the same semimajor axis as the Lagrange point
of the Earth but varying in longitude. The test particles surviving after
1, 5, 10 and 25 million years are shown in green, blue, yellow and
red respectively.}
\label{fig:earthinclineplotcol}
\end{figure*}
\begin{figure*}
\centerline{\psfig{figure=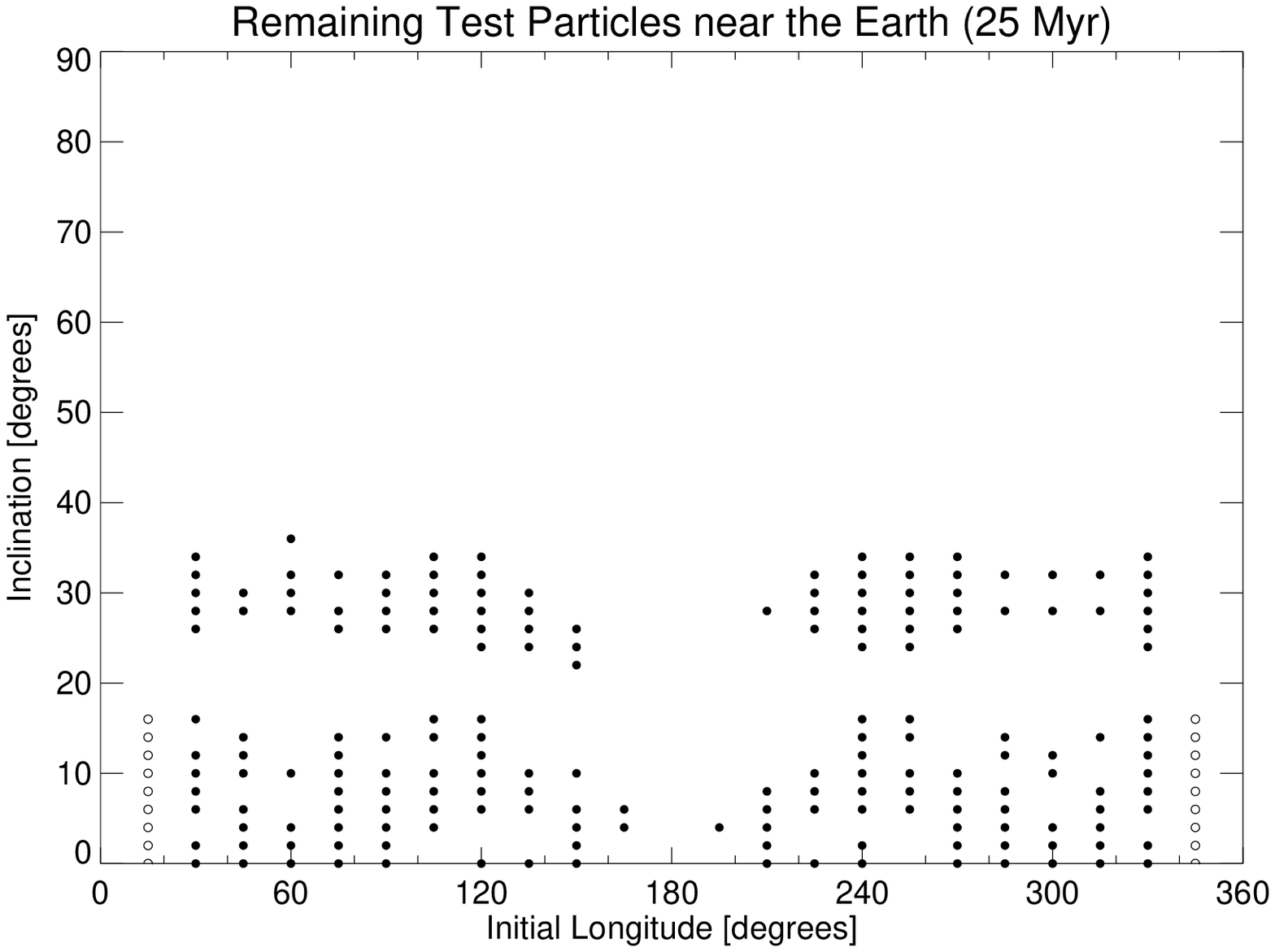,height=0.4\hsize}}
\caption{This shows a close-up of the stability zones of the inclined
survey of terrestrial test particles. Only those surviving the full
integration time of 25 million years are plotted. The filled circles
are tadpole orbits, while the open circles are horseshoe orbits. There
seem to be two bands of stability, one at low starting inclinations
($i \lta 16^\circ$) and one at moderate starting inclinations
($24^\circ \lta i \lta 34^\circ$).  }
\label{fig:earthinclineplotbw}
\end{figure*}

\section{Venusian Surveys}

Perhaps one of the likeliest planets in the inner Solar System to
harbour undiscovered Trojans is Venus.  Fig.~\ref{fig:venusplaneplot}
shows the results of the in-plane survey of Venusian test
particles. The starting semimajor axes are scaled by a fraction of the
planet's semimajor axis in steps of $0.0012$, while the argument of
pericentre is offset from that of the planet by $5^\circ$ steps. The
test particles are integrated for 25 million years and the survivors
recorded in the Figure.  A filled circle represents a tadpole orbit,
an open circle represents a horseshoe orbit.  Notice that there are
some long-lived survivors on horseshoe orbits even around the
conjunction point. The tadpole orbits survive around $\Lfour$ for
starting longitudes between $15^\circ$ and $160^\circ$ and around
$\Lfive$ for starting longitudes between $195^\circ$ and
$345^\circ$. The offset in the semimajor axes of the survivors $\Delta
a$ compared to the parent planet $\aplanet$ satisfy $\Delta a/
\aplanet \lta 0.72 \%$.  There are 792 test particles at the beginning
of the simulation, but only 407 persist till the end, of which 168 are
on tadpole orbits.  Horseshoes and tadpoles are of course divided by a
separatrix in phase space (see Section 2). The break-up of the
separatrix is associated with a chaotic layer, and this is responsible
for erosion between the filled and open circles in the stable zones.
At the edge of the figure, we show the stability boundary inferred
from the elliptic restricted three body problem.  The diamonds
represent unstable test particles in the elliptic restricted three
body problem (comprising the Sun, Venus and the massless
asteroid). Everything within this outer boundary of diamonds is stable
at the level of the restricted three body
problem. Fig.~\ref{fig:venusextremaplot} shows an at first sight
surprising regularity of the orbits of the survivors. The extrema of
the semimajor axis $\aext$ of the Trojan test particles are plotted
against the initial longitude and fall on a one-parameter family of
curves according to the semimajor axis factor. The heliocentric
Hamiltonian for a test particle in the frame rotating with the mean
motion of the planet may be written
\begin{equation}
H = - {k^2\over 2a} - \mplanet k^2 R - \nplanet \sqrt{(1 +\mplanet)a}.
\end{equation}
Here, $\nplanet = \sqrt{(1 + \mplanet)/\aplanet^3}$ is the mean motion
of the planet using Kepler's third law, and $R$ is approximated by the
zeroth order term in the disturbing function
\begin{equation}
R = {1\over \aplanet} \left[ {1\over 2| \sin \half \phi|} -\cos
\phi \right],
\end{equation}
where $\phi$ is the difference in longitude between the planet and the
asteroid. This follows on setting $a = \aplanet$ in
eq.~(\ref{eq:zeroth}). This Hamiltonian depends on time only through the slow
variation of the planet's orbital elements. These take place on a
timescale much longer than the libration of the Trojan, and so the
Hamiltonian is effectively constant. Expanding in the difference
between the semimajor axis of the planet and the Trojan, we readily
deduce that a test particle with a semimajor axis factor $f$ and which
starts at a differential longitude $\theta$ has extrema $\aext$
satisfying
\begin{equation}
{\aext \over \aplanet} = 1 + \left[ (f-1)^2 + {8\mu\over 3}
\left({1\over 2 |\sin \half \theta |} - \cos \theta -
{1 \over 2} \right) \right]^\half,
\label{eq:extrema}
\end{equation}
where $\mu = \mplanet/(1 + \mplanet)$ is the reduced mass.  The
extrema of the semimajor axis of the Trojans fall on this
one-parameter family of curves, as depicted in
Fig.~\ref{fig:venusextremaplot}.

Fig.~\ref{fig:venuspanelsplot} shows distributions of the
eccentricities and inclinations of the survivors plotted against their
initial semimajor axis and longitude from Venus. In all the panels,
the maximum values attained during the course of the 25 million year
orbit integrations are marked by crosses, the instantaneous values are
marked by circles. There are two striking features of these diagrams.
First, the eccentricities and the inclinations of the survivors remain
very low indeed. The mean eccentricity of the sample after 25 million
years is $0.027$, while the mean inclination is $0.706^\circ$.  This
is a very stable family. Second, most of the survivors seem to occupy
very nearly the same regions of the plots. The striking horizontal
line of crosses in the rightmost panels of
Fig.~\ref{fig:venuspanelsplot} suggest that these test particles do
lie on similar orbits belonging to similar families exploring similar
regions of phase space.

Fig.~\ref{fig:venusinclineplotcol} shows the results of the Venusian
inclined survey. Here, the orbits of 1104 test particles around Venus
are integrated for 75 million years. The initial inclinations of the
test particles (with respect to the plane of Venus' orbit) are spaced
every $2^\circ$ and the initial longitudes (again with respect to
Venus) are spaced every $15^\circ$. The test particles are
colour-coded according to whether they survive till the end of the 5,
20, 50 and 75 million year integration timespans.  The 137 survivors
after 75 million years all have smallish inclinations ($i <
16^\circ$). Notice, too, that the stable zones around the Lagrange
points are still connected by some surviving test particles at
conjunction. Bodies trapped around the Lagrange points should be
sought at all longitudes close to the orbital
plane. Fig.~\ref{fig:venusinclineplotbw} shows a close-up of the
stable zones, with tadpole orbits represented by filled circles and
horseshoe orbits by open circles. Most of the objects that do survive
are true Trojans, in that there orbits are recognisably of a tadpole
character. There are just 8 surviving horseshoe orbits. The ensemble
has a mean eccentricity of $0.041$ and a mean inclination of
$6.941^\circ$.

Assuming that they are primordial, we can estimate the number of
coorbiting Venusian satellites by extrapolating from the number of
Main Belt asteroids (c.f., Holman 1997, Evans \& Tabachnik 1999). The
number of Main Belt asteroids $N_{\rm MB}$ is $N_{\rm MB} \lta
\Sigma_{\rm MB} A_{\rm MB} f$, where $A_{\rm MB}$ is the area of the
Main Belt, $\Sigma_{\rm MB}$ is the surface density of the
proto-planetary disk and $f$ is the fraction of primordial objects
that survive ejection (which we assume to be a universal
constant). Let us take the Main Belt to be centred on $2.75$ AU with a
width of $1.5$ AU.  Fig.~\ref{fig:venusplaneplot} suggests that the
belt of Venusian Trojans is centred on $0.723$ AU and has a width of $
\lta 0.008$ AU. If the primordial surface density falls off inversely
proportional to distance, then the number of coorbiting Venusian
asteroids $N_{\rm V}$ is
\begin{equation}
N_{\rm V} \lta \Bigl( {2.75\over 0.723} \Bigr) \Bigl( {0.723 \times 0.008
\over 2.75 \times 1.5} \Bigr)  N_{\rm MB}
\approx 0.0053 N_{\rm MB}.
\label{eq:approxvenustrojans}
\end{equation}
The number of Main Belt asteroids with diameters $\gta 1$ km is $\sim
40000$, which suggests that the number of Venusian Trojans is $\sim
100$ with perhaps a further $\sim 100$ coorbiting companions on
horseshoe orbits.

\begin{figure*}
\centerline{\psfig{figure=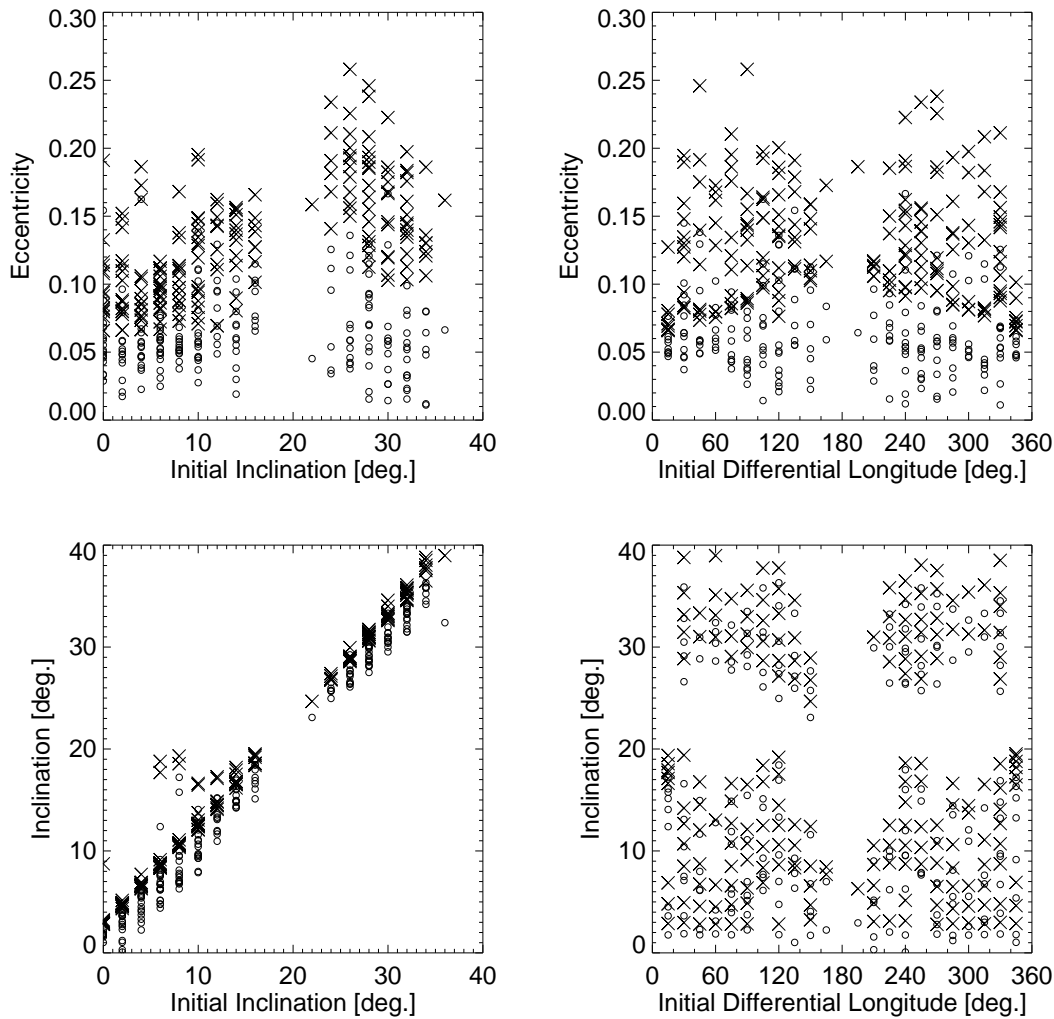,height=0.8\hsize}}
\caption{This shows the distributions of eccentricities and
inclinations of the surviving test particles of
Fig.~\ref{fig:earthinclineplotbw} against their initial semimajor axis
and longitude from the Earth. In all the panels, the maximum values
attained during the course of the orbit integrations are marked by
crosses, the instantaneous values after 25 million years are marked by
circles.  The bimodality of the inclination distribution of the
survivors is manifest.}
\label{fig:earthpanelsplot}
\end{figure*}
\begin{figure*}
\centerline{\psfig{figure=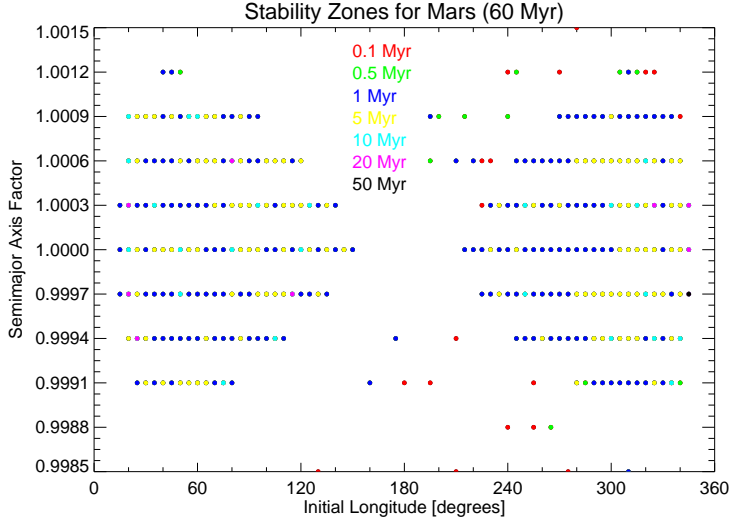,height=0.4\hsize}}
\caption{This shows the erosion of the in-plane test particles near
Mars. There are no surviving test particles after 60 million years.
The test particles are colour-coded according to their survival times.}
\label{fig:marsinplane}
\end{figure*}
\begin{figure*}
\centerline{\psfig{figure=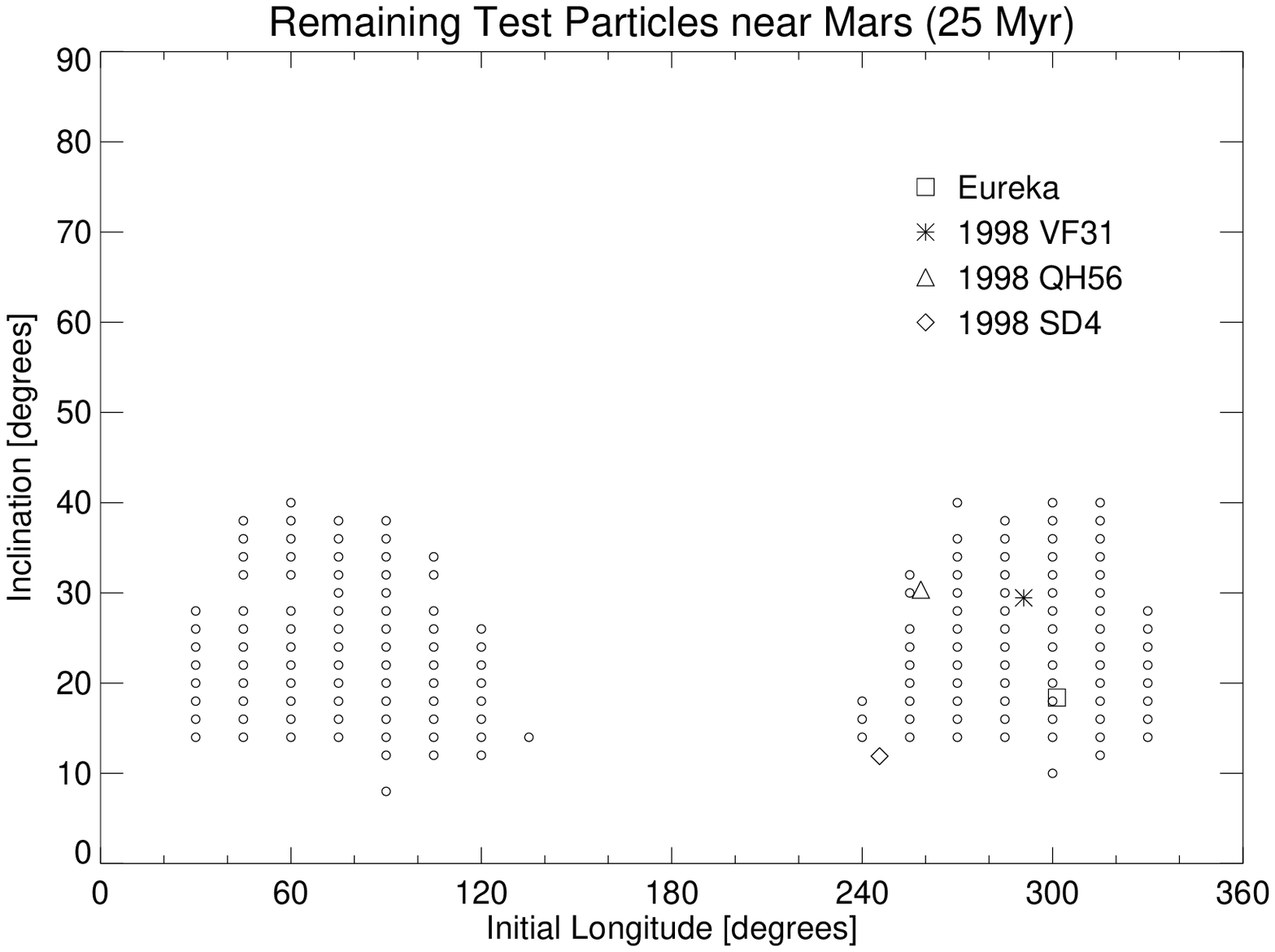,height=0.4\hsize}}
\caption{This shows the results of the survey of inclined test
particles near Mars. The horizontal axis marks the longitude measured
from Mars and the vertical axis the inclination with respect to Mars
of the starting positions of test particles.  Only the particles
surviving till the end of the 25 million year integration are
marked. Also shown are the instantaneous positions of the two Martian
Trojans, namely 5261 Eureka and 1998 VF31, as well as the asteroids
1998 QH56 and 1998 SD4. The latter two have been suggested as possible
Trojans, though this now seems unlikely.}
\label{fig:marsinclineplot}
\end{figure*}

\section{Terrestrial Surveys}

The Earth is slightly more massive than Venus, and this augurs well
for the existence of coorbiting satellite companions. The Earth has
no known Trojans (Whiteley \& Tholen 1998), but it does possess the
asteroidal companion 3753 Cruithne which moves on a temporary
horseshoe orbit (Wiegert, Innanen \& Mikkola 1997; Namouni, Christou
\& Murray 1999). The asteroid will persist on this horseshoe orbit for
a few thousand years.

Fig.~\ref{fig:earthplaneplot} shows the surviving in-plane test
particles near the Earth after their orbits have been integrated for
50 million years.  Again, filled circles represent the tadpole orbits,
open circles the horseshoe orbits.  On comparison with
Fig.~\ref{fig:venusplaneplot}, we see that the stable zones of the
Earth are more extensive and the number of survivors is greater than
for Venus. Tadpole orbits survive for $\Delta a/\aplanet \lta 0.48 \%$
and horseshoes for $\Delta a/\aplanet \lta 1.20 \%$. Although the
number of survivors is greater, the number of true Trojans on tadpole
orbits is much less than for Venus. Specifically, of the 792 original
test particles, only 509 persist till the end of the simulation and of
these just 95 are on tadpole orbits. In the case of the Earth, just
$19 \%$ of the survivors are tadpoles, as opposed to $41 \%$ for
Venus. The visual consequence of this is that the holes in the stable
regions for the Earth are more pronounced than for Venus (compare also
the equivalent diagram for Saturn provided by Holman \& Wisdom
1993). The survivors have a mean eccentricity of $0.038$ and a mean
inclination of $1.349^\circ$, consistent with a long-lived population.

Fig.~\ref{fig:earthinclineplotcol} shows the erosion of an ensemble of
1104 inclined test particles positioned at the same semimajor axis as
the Earth but varying in longitude.  Again, the the initial
inclinations of the test particles (with respect to the plane of the
Earth's orbit) are spaced every $2^\circ$ and the initial longitudes
are spaced every $15^\circ$. The test particles are colour-coded
according to their survival times -- those surviving after 1, 5, 10
and 25 million years are shown in green, blue, yellow and red
respectively. The ensemble after 25 million years is shown in
Fig.~\ref{fig:earthinclineplotbw} with tadpoles represented as filled
circles and horseshoes as open circles. Again, the first thing to note
is the number of survivors -- there are 200 in total, almost all of
which are moving on tadpole orbits.  There seem to be two bands of
stability, one at low starting inclinations ($i \lta 16^\circ$) and
one at moderate starting inclinations ($24^\circ \lta i \lta
34^\circ$). On careful inspection of our earlier
Fig.~\ref{fig:venusinclineplotcol}, it is possible to discern in blue
and green a similar band of stable trajectories at moderate
inclination for Venus. These though are swept out much more quickly
than for the Earth; they are all gone after just 5 million years.  The
survivors in Fig.~\ref{fig:earthinclineplotbw} have a low mean
eccentricity of $0.064$. The distribution of inclinations is
strikingly bimodal as is evident from
Fig.~\ref{fig:earthpanelsplot}. The mean inclination at the end of the
simulation is $15.778^\circ$.

Let us remark that the Earth has more surviving test particles in both
of our surveys than Venus, as well as more extensive stable
zones. This suggests that the asteroid 3753 Cruithne may not be
unique, but the first member of a larger class of coorbiting
terrestrial companions (Namouni, Christou \& Murray 1999). Making the
same assumptions as in~(\ref{eq:approxvenustrojans}), our estimate for
the number of terrestrial companions is
\begin{equation}
N_{\rm E} \lta \Bigl( {2.75\over 1.00} \Bigr) \Bigl( {1.00 \times 0.01
\over 2.75 \times 1.5} \Bigr)  N_{\rm MB}
\approx 260.
\label{eq:approxearthtrojans}
\end{equation}
Here, we have used Fig.~\ref{fig:earthplaneplot} to set the width of
the stable zone around the Earth as $\sim 0.005$ AU. The total number
of coorbiting companions will be much higher, if one includes
transient objects like 3753 Cruithne.
\begin{figure*}
\centerline{\psfig{figure=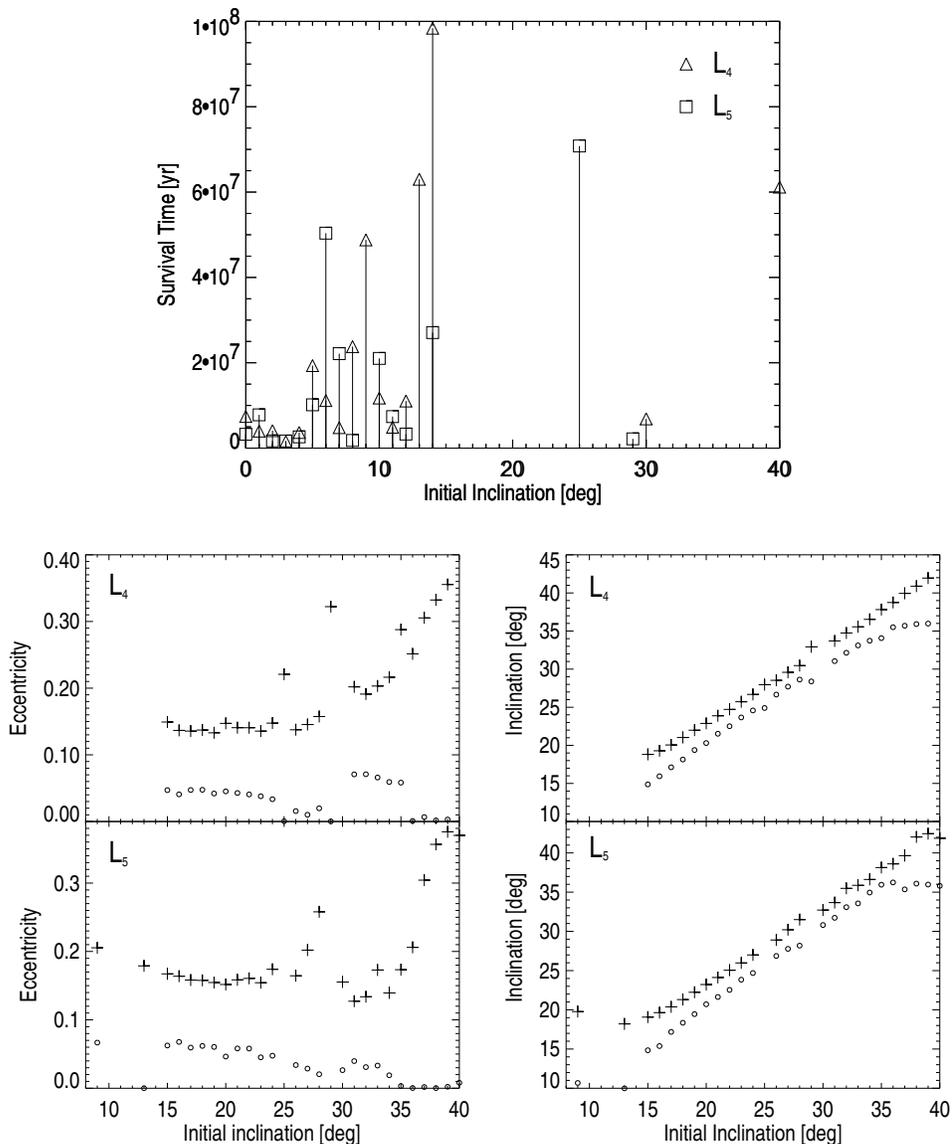,height=0.8\hsize}}
\vskip 0.5truecm
\caption{This upper panel shows the termination time of the 100 Myr
survey of the Martian Lagrange points. Test particles at $\Lfour$ are
identified by triangles while squares mark test particles at
$\Lfive$. The instantaneous (open circles) and the maximum (crosses)
values of the eccentricity and inclination are illustrated in the four
lower panels.}
\label{fig:marsprinceton}
\end{figure*}

\begin{figure*}
\centerline{\psfig{figure=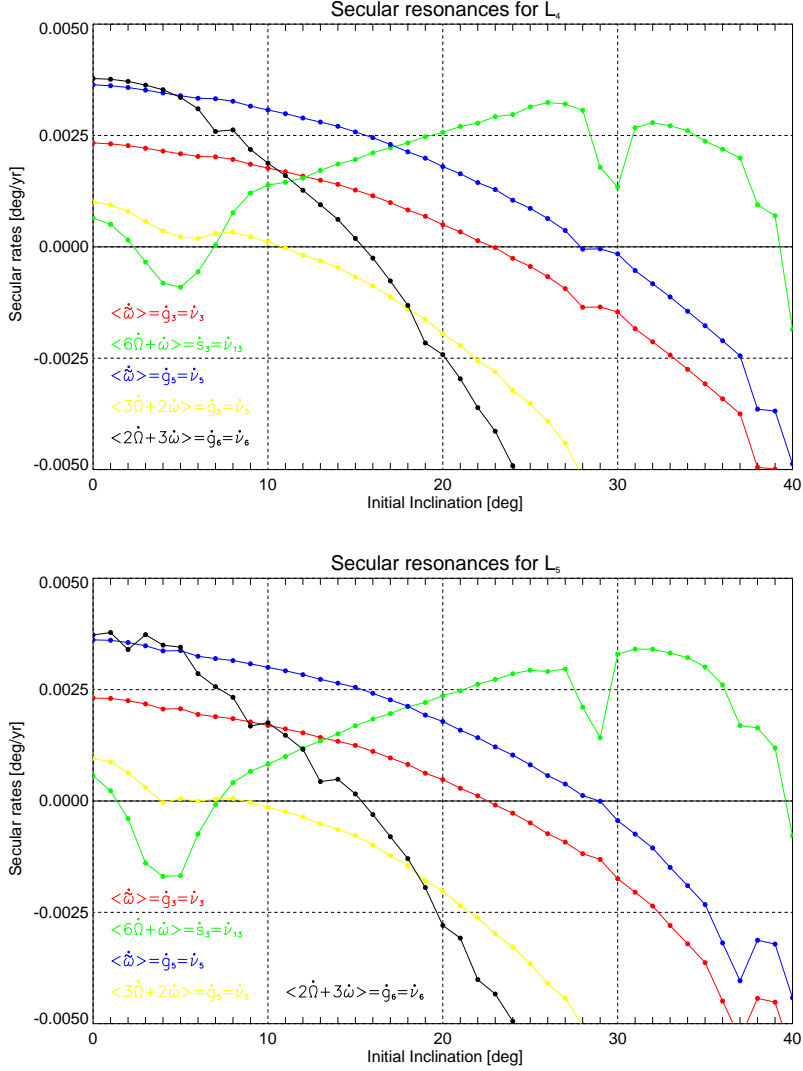,height=0.75\hsize}}
\vspace{0.5cm}
\caption{This shows the locations of the principal secular resonances
for the $\Lfour$ and $\Lfive$ points of Mars. The rates of change of
the angles (in degrees per year) are plotted against inclination.
Resonances occur when these curves cross the horizontal axis of zero
rate of change. Notice the strong linear resonance with Jupiter (blue
curve). This occurs at $\sim 28-30^\circ$ at the $\Lfour$ point and at
$\sim 29^\circ$ at the $\Lfive$ point. Many of the low inclination
test particles are destabilised by the non-linear Jovian resonance
(yellow curve), which persists over a range of inclinations between
$\sim 6-11^\circ$ . There are weaker resonances with the Earth (red
and green curves) as well as Saturn (black curve).  }
\label{fig:marsresonances}
\end{figure*}
\begin{figure*}
\centerline{\psfig{figure=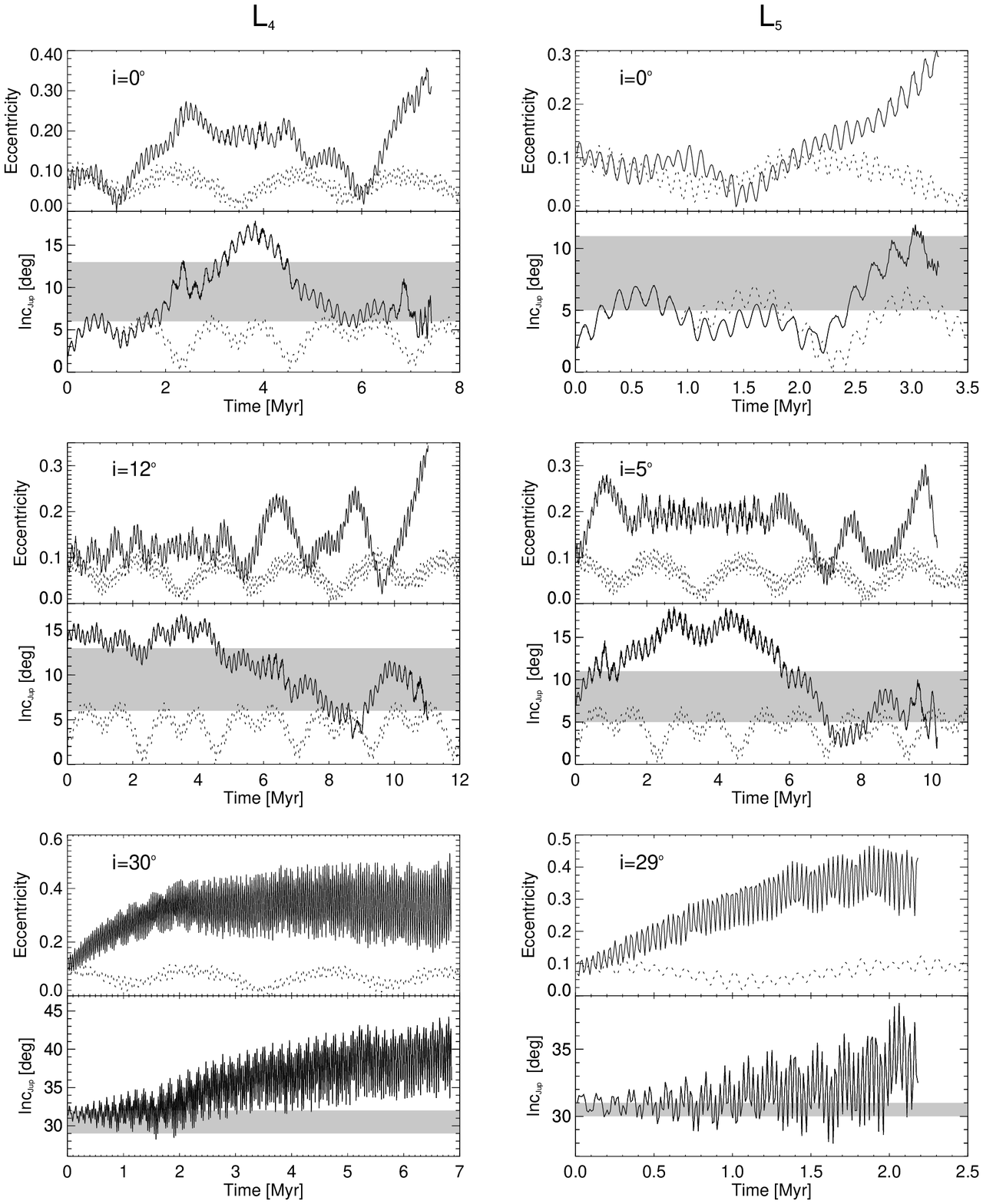,height=\hsize}}
\caption{This shows the evolution of the inclination (with respect to
the plane of Jupiter) and the eccentricity of selected orbits in full
lines. The behaviour of Mars is shown in a broken line.  The
inclinations over which the two Jovian resonances operate are shown as
shaded grey bands. The left column refers to $\Lfour$, the right
column to $\Lfive$. In each case, the starting inclination with respect 
to the plane of Mars is noted in the panel.
}
\label{fig:marsorbits}
\end{figure*}

\section{Martian Surveys}
Mars is the only terrestrial planet already known to possess Trojan
asteroids. These are 5261 Eureka and 1998 VF31 (see, for example,
Mikkola et al. 1994, Tabachnik \& Evans 1999 and the references
therein). Both have moderate inclinations to the ecliptic, namely
$20.3^\circ$ and $31.3^\circ$ respectively.

The in-plane Martian survey is presented in
Fig.~\ref{fig:marsinplane}. The test particles are colour-coded
according to their survival times. Most of the particles are already
swept out after 5 million years. After 50 million years, there is only
one test particle remaining, and it too is removed shortly
afterwards. There are no survivors after 60 million years. This
confirms the earlier suspicions of Mikkola \& Innanen (1994) that
Trojans in the orbital plane of Mars are not long-lived.  As is
evident from Table~\ref{table:statisticsinplane}, the most common fate
of the test particles is to enter the sphere of influence of Mars.
The inclined survey is shown in Fig.~\ref{fig:marsinclineplot},
together with the positions of the two Trojans, 5261 Eureka (marked by
a square) and 1998 VF31 (asterisk). Two further asteroids -- 1998
QH56 (triangle) and 1998 SD4 (diamond) -- have been suggested
as Trojan candidates, although improved orbital elements together with
detailed numerical simulations (Tabachnik \& Evans 1999) now make this
seem rather unlikely. The result of the inclined survey is to show
stable zones for inclinations between $14^\circ$ and $40^\circ$ for
timespans of 25 million years. The stable zones are strongly eroded at
$\sim 29^\circ$.

Following this, we conducted another experiment in which inclined
Martian Trojans are simulated for 100 Myrs. The initial conditions are
inherited from Mars, except for the argument of pericentre which is
offset by $60^\circ$ ($\Lfour$) and $300^\circ$ ($\Lfive$), and the
inclinations which are selected in the range from $0^\circ - 40^\circ$
from Mars' orbital plane. To examine in more detail the effects of the
other planet's perturbations, a 1-degree step in inclination is
chosen. Fig.~\ref{fig:marsprinceton} shows the results of this
exercise. The upper panel expresses the termination time versus the
initial inclinations of the test particles at both Lagrange points.
Not surprisingly, low inclination Trojans ($i < 5^\circ$) enter Mars'
sphere of influence on a 10 million year timescale. The stable
inclination windows are also recovered with a strong disturbing
mechanism at $29^\circ$ for $\Lfive$ and $30^\circ$ for $\Lfour$. The
four lower panels give the eccentricities and inclinations of the
remaining test particles at the end of the integration. Crosses
identify the maximum quantities over the entire timespan, while open
circles show the instantaneous values at 100 Myr. The general trend is
to have stable orbits ($e_{\rm max} < 0.2$) in the range $15^\circ
-34^\circ$ in the case of $\Lfour$ and $9^\circ - 36^\circ$ in the
case of $\Lfive$. Interestingly, the two securely known Trojans,
namely 5261 Eureka and 1998 VF31, occupy positions at $\Lfive$
corresponding to the two local minima of the maximum eccentricity
curve.

The results for the inclined Martian Trojans are unusual, and it is
natural to seek an explanation in terms of secular resonances.  Large
disturbances can occur when there is a secular resonance, that is,
when the averaged precession frequency of the asteroid's longitude of
pericentre $\avvarpidot$ or longitude of node $\avOmegadot$ becomes
nearly equal to an eigenfrequency of the planetary system (e.g.,
Brouwer \& Clemence 1961; Williams \& Faulkner 1981; Scholl et
al. 1989). The secular precession frequencies in linear theory are
usually labelled $g_j$ and $s_j$ ($j= 1, \dots 8$ for Mercury to
Neptune) for the longitude of pericentre and longitude of node
respectively. Their mean values computed over 200 million years are
listed in Laskar (1990).

Fig.~\ref{fig:marsresonances} can be used to infer the positions of
some of the principal secular resonances as a function of inclination
in the vicinity of each Lagrange point. The vertical axis is the rate
of variation of various angles, the horizontal axis is the
inclination. A resonance occurs whenever the rate of variation
vanishes.  The blue curve shows the frequency $\avvarpidot - g_5$,
which vanishes at inclinations $\sim 28-30^\circ$ for the $\Lfour$
point and $29^\circ$ at the $\Lfive$ point.  Notice that the resonance
is much broader at the $\Lfour$ point.  The yellow curves show
$\langle 3\Omegadot + 2\omegadot \rangle - g_5$. This frequency
vanishes at a range of inclinations between $\sim 6-11^\circ$, again
with slight differences noticeable at the two Lagrange points. As
these are resonances with Jupiter, they are expected to be the most
substantial.  At this semimajor axis, there is just one resonance with
Saturn. The black curves show the frequency $\langle 2\Omegadot +
3\omegadot \rangle - g_6$. This resonance occurs at inclinations of
$\sim 15^\circ$. Lastly, there are two weaker resonances with the
Earth. These may be tracked down using the red curve, which shows
$\avvarpidot - g_3$, and the green curve, which shows $\langle
6\Omegadot + \omegadot \rangle -s_3$. This completes the list of the
main resonances in the vicinity of Mars. The importance of the Jovian
resonances in particular has been pointed out before by Mikkola \&
Innanen (1994). 
 
Fig.~\ref{fig:marsorbits} illustrates the evolution of a few
arbitrarily selected orbits. The left panels refer to the $\Lfour$
Lagrange point, the right panels to the $\Lfive$ point. The panels are
labelled according to the initial inclination with respect to Mars'
orbit. They plot the evolution of the eccentricity $e$ and the
inclination with respect to Jupiter $\ijup$ for a typical test
particle (full curve) and Mars (broken curve). The inclinations over
which the two Jovian resonances operate are shown as shaded bands.
The uppermost two panels refer to orbits that start out in the orbital
plane of Mars. Their inclination is initially increased, and this
takes the orbits into the r\'egime in which one of the Jovian
resonances is dominant. The eccentricity of the orbit is pumped
whenever it lingers in the shaded inclination band. This makes the
orbit Mars-crossing and the test particle is terminated. The middle two
panels show the fates of orbits at intermediate inclinations of
$12^\circ$ at the $\Lfour$ point and $5^\circ$ at the $\Lfive$
point. Although the behaviour is quite complex, the final increase in
eccentricity in both cases coincides with prolonged stays in the
resonant region.  We conclude that the low inclinations test particles
are destabilised by this secular resonance with Jupiter. The bottom
two panels show fates of test particles starting off at $30^\circ$ at
the $\Lfour$ point and $29^\circ$ at the $\Lfive$ point. In both
cases, there is a rapid and pronounced increase in the eccentricity,
which takes it onto a Mars-crossing path. This destabilisation occurs
only for a very narrow range of inclinations.  This is manifest in the
erosion in Fig.~\ref{fig:marsinclineplot}, especially at the $\Lfour$
point near inclinations of $30^\circ$.

Mikkola \& Innanen (1994) suggested that the instability of low
inclination Martian Trojans was due to a secular resonance with Mars
driving the inclination upwards. In their picture, this continues
until a critical inclination of $\sim 12^\circ$ is reached when
$3\Omega + 2\omega$ resonates with Jupiter. The difficulty with this
is that it is not clear whether the claimed Martian secular resonance
-- which is really just equivalent to the statement that the test
particle is coorbiting -- is responsible for the inclination increase.
Our Fig.~\ref{fig:marsresonances} seems to show that $\langle
3\Omegadot + 2\omegadot \rangle - g_5$ nearly vanishes over a broad
range of inclinations and we suspect it may be able to cause the
damage on its own.

\begin{figure}
\centerline{\psfig{figure=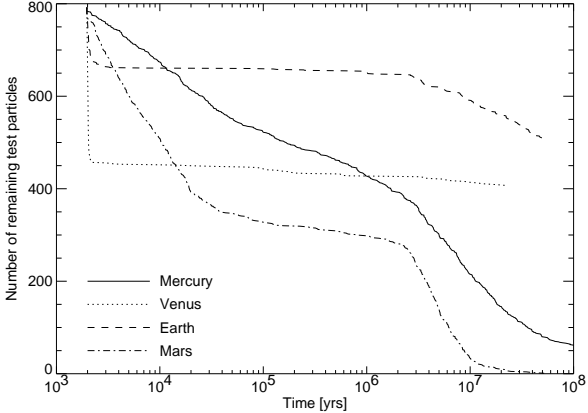,height=0.7\hsize}}
\caption{The number of remaining test particles is plotted against
time for our in-plane surveys of the Lagrange points of the four
terrestrial planets. The extrapolations presented in
Table~\ref{table:predictions} suggest that there will be some hundreds
of surviving test particles for Venus and the Earth even if the
simulations are continued for the age of the Solar System ($\sim 5$
Gyrs).}
\label{fig:ntinplane}
\end{figure}
\begin{table*}
\begin{center}
\begin{tabular}{rccccc}
Planet & $a$ & $b$ & $N(3000 \yrs)$ & $N_{\rm exp}$ (1Gyr) & $N_{\rm exp}$
(5Gyr)\\ \hline
Mercury & $1283.2\pm 5.3$ &$-150.6 \pm 0.89$ & $756$ & $-$ & $-$ \\
Venus & $503.5\pm 1.9$ & $ -12.61 \pm 0.33$ & $456$ & $390$ &$381$ \\
Earth & $946.8\pm 18.8$ & $-52.36 \pm 2.71$ & $663$ & $476$ & $439$ \\
Mars & $1089.6\pm 8.2$ & $-142.33 \pm 1.53$ & $691$ &$-$& $-$ \\
\null&\null&\null&\null&\null&\null \\
\end{tabular}
\end{center}
\caption{For each planet, the data on the number of surviving in-plane
test particles $N(t)$ is fitted for $t >3000$ yr.  The parameters $a$
and $b$ in the logarithmic fits are listed in the second and third
columns. The last two columns of the tables give the extrapolated
number of test particles estimated to remain after 1 Gyr and 5 Gyr.}
\label{table:predictions}
\end{table*}
\begin{figure}
\centerline{\psfig{figure=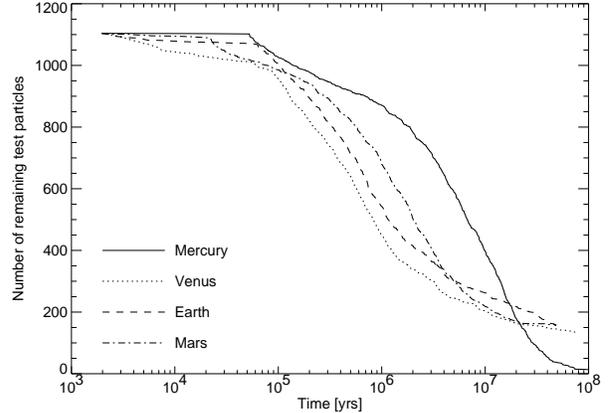,height=0.7\hsize}}
\caption{As Fig.~\ref{fig:ntinplane}, but for our surveys of inclined
orbits near the Lagrange points of the four terrestrial planets. The
decay laws for test particles around all four planets are similar and
they are not well-fit by a logarithmic decay law.}
\label{fig:ntinclined}
\end{figure}
\begin{table*}
\begin{center}
\begin{tabular}{rccccc}
Planet & $c$ & $d$ & $N$ (10 Myr) & $N_{\rm exp}$ (1Gyr) & $N_{\rm exp}$
(5Gyr)\\ \hline
Mercury & $13.89 \pm 0.07$ &$1.60 \pm 0.01$ & $396$ & $-$ & $-$ \\
Venus & $3.62 \pm 0.03$ & $0.19 \pm 0.005$ & $203$ & $80$ &$59$ \\
Earth & $4.61 \pm 0.03$ & $0.31 \pm 0.005$ & $262$ & $63$ & $38$ \\
Mars & $4.037 \pm 0.1$ & $0.25 \pm 0.014$ & $217$ &$67$& $45$ \\
\null&\null&\null&\null&\null&\null \\
\end{tabular}
\end{center}
\caption{For each planet, the data on the number of surviving inclined
test particles $N(t)$ is fitted for $10 < t < 100$ Myr.  The parameters $c$
and $d$ in the power-law fits are listed in the second and third
columns. The last two columns of the tables give the extrapolated
number of test particles estimated to remain after 1 Gyr and 5 Gyr.}
\label{table:morepredictions}
\end{table*}

\section{Conclusions}

The possible existence of long-lived coorbiting satellites of the
terrestrial planets has been examined using numerical simulations of
the Solar System.  Of course, integrations in the inner Solar System
are laborious as much smaller timesteps are required to follow the
orbits of the satellites of Mercury as opposed to the giant planets
like Jupiter.  Our numerical surveys have been pursued for timescales
up to 100 million years -- typically an order of magnitude greater
than previous computations in the inner Solar System.  The numerical
algorithm is a symplectic integrator with individual timesteps that
incorporates the most important post-Newtonian corrections (Wisdom \&
Holman 1991; Saha \& Tremaine 1994).

One worry concerning our integrations is that they extend at most to
100 million years, which is a small fraction of the age of the Solar
System ($\sim 5$ Gyrs). To gain an idea of the possible effects of
longer integration times, we can use the approximate device of fitting
our existing data and extrapolating.  Graphs of the number of
surviving particles against time are shown in
Figs.~\ref{fig:ntinplane} -~\ref{fig:ntinclined} for the in-plane and
inclined surveys respectively.  In the former case, the data are
generally well-fit by a logarithmic decay law of form
\begin{equation}
N(t) = a + b \log_{10} \Bigl( t [\yrs] \Bigr)
\end{equation}
Table~\ref{table:predictions} shows the best fitting values of $a$ and
$b$. It also gives the extrapolated number of test particles after 1
and 5 Gyrs. For both Venus and the Earth, this suggests that several
hundred test particles remain, even if the simulations are run for the
age of the Solar System. In the inclined case, the data is not
well-fitted by logarithmic decay laws. Instead, 
Table~\ref{table:morepredictions} gives the results of fitting the
data between 10 and 100 Myr to a power-law decay of form
\begin{equation}
N(t) = {10^c \over \Bigl( t [\yrs] \Bigr)^d}
\end{equation}
Extrapolation suggests that some inclined test particles remain for
Venus, the Earth and Mars even after 5 Gyr. However, these numbers
are highly speculative for the inclined survey, as only the tail of
the distribution is fitted.

The results of our surveys for Venus and the Earth are somewhat
similar. Long-lived coorbiting satellites can persist in the orbital
planes of both planets. The stable zones of the Earth are larger than
those Venus, although the Earth retains fewer true Trojans or
tadpoles. The semimajor axes of the stable test particles $a$, as
compared to the parent planet $\aplanet$, satisfy $\Delta a/ \aplanet
\lta 0.72 \%$ for Venus and  $\Delta a/ \aplanet \lta 1.2 \%$ for the 
Earth. Both Venus and the Earth have low inclination r\'egimes in
which long-lived test particles survive for timescales of tens of
millions of years, despite the disturbing perturbations from the
remainder of the Solar System.  The stable zones satisfy $i \lta
16^\circ$ for Venus. For the Earth, there are two bands of stability,
one at low inclinations ($i \lta 16^\circ$) and one at moderate
inclinations ($24^\circ \lta i \lta 34^\circ$).  However, there is a
hint that the higher inclination band may be further eroded with still
longer timespan integrations.  The inclined test particles that
survive primarily move on tadpole orbits. It is possible to make very
crude estimates of numbers by extrapolation from the Main Belt. These
suggest that there may be some hundreds of asteroids in the coorbital
regions of Venus and the Earth.

In the case of Mercury, very few of the test particles -- whether
starting in the orbital plane or at higher inclinations -- survive
till the end of the integrations. This seems reasonable, as both the
low mass and the high eccentricity of Mercury militate against stable
zones. A population of numerous, long-lived Mercurian Trojans seems
rather unlikely. Recently, Namouni et al. (1999) have speculated that
highly inclined coorbiting Mercurian satellites (``Vulcanoids'') may
exist. Our survey does hint at the survival of a handful of test
particles at high inclinations. On re-simulating these results over
the same timespan using an updated ephemerides, these test particles
did not survive but were ejected. So, if there are stable zones of
inclined coorbiting satellites, then they must be narrow and rather
sensitive to the detailed initial conditions.

From the viewpoint of dynamics, Mars offers perhaps the greatest
interest of all. Here, the test particles within the orbital plane are
all ejected on 60 million year timescales.  The inclined survey shows
stable zones for inclinations between $14^\circ$ and $40^\circ$.  This
is certainly consistent with the two known Martian Trojans (5261
Eureka and 1998 VF31) which have orbits moderately inclined to the
ecliptic ($20.3^\circ$ and $31.3^\circ$ respectively) about the
$\Lfive$ point.  The survey shows strong erosion at a narrow band of
inclinations concentrated around $\sim 28-30^\circ$ at $\Lfour$ and
$29^\circ$ at $\Lfive$.  This may be traced to a strong, but narrow,
Jovian resonance. The averaged precession frequency of the test
particle's longitude of pericentre $\avvarpidot$ is equal to the
secular precession frequency of the longitude of pericentre of Jupiter
$g_5$. We believe that the destabilisation of the low inclination and
in-plane orbits is also due to Jupiter. Here, the combination of
frequencies $\langle 3\Omegadot + 2\omegadot \rangle - g_5$ vanishes
over a broad range of inclinations between $\sim 6-11^\circ$. As they
evolve, low inclination test particles enter this band of
inclinations, their eccentricity is increased and they become
Mars-crossing.

\section*{Acknowledgments}
NWE is supported by the Royal Society, while ST acknowledges financial
help from the European Community.  We wish to thank John Chambers,
Jane Luu, Seppo Mikkola, Fathi Namouni, Prasenjit Saha and Scott
Tremaine for helpful comments and suggestions. Jane Luu and Prasenjit
Saha provided critical readings of the manuscript.

\begin{appendix}

\section{Auxiliary Formulae}

This appendix lists auxiliary functions for the averaged disturbing
function
\begin{eqnarray}
U_{2,0} & =& {a\over \aplanet^2} \Bigl[
\half (e^2 + \eplanet^2)\cos \phi - e \eplanet \cos(\omega +
\Omega + 2\phi)\Bigr]\cos^2 \halfi , \nonumber \\
U_{2,3} & =& \Bigl[ \ffrac{1}{4}e\eplanet a\aplanet \cos
(\omega+\Omega+2\phi) - \half(e^2+\eplanet^2)a\aplanet\cos\phi \nonumber \\
& & +\ffrac{9}{4}e\eplanet a \aplanet\cos(\omega + \Omega) \Bigr]
\cos^2 \halfi - \ffrac{3}{4}(a^2e^2 + \aplanet^2\eplanet^2),\nonumber
\\
U_{2,5} & =& \Bigl[\ffrac{15}{8}(e^2 + \eplanet^2)a^2 \aplanet^2 
+\ffrac{27}{8}e\eplanet a^2\aplanet^2\cos(\omega+\Omega-\phi) \nonumber \\
& & -\ffrac{9}{4} e\eplanet a^2\aplanet^2\cos (\omega + \Omega + \phi) 
-\ffrac{9}{8}(e^2+\eplanet^2)a^2 \aplanet^2\cos 2\phi \nonumber \\
& & + \ffrac{3}{8}e \eplanet a^2 \aplanet^2
\cos (\omega + \Omega + 3\phi)\Bigr] \cos^4 \halfi + \nonumber \\
& & \Bigl[ \ffrac{3}{4} e\eplanet a\aplanet (a^2+\aplanet^2) 
\cos(\omega + \Omega + 2\phi) - \ffrac{3}{2}a\aplanet 
(a^2e^2+ \nonumber\\
& & \aplanet^2\eplanet^2)\cos\phi - \ffrac{9}{4}e\eplanet 
a\aplanet(a^2 + \aplanet^2)\cos(\omega + \Omega)\Bigr]\cos^2\halfi
\nonumber \\
& & + \ffrac{3}{4}(a^4 e^2 + \aplanet^4 \eplanet^2) +
\ffrac{3}{2}e\eplanet a^2\aplanet^2 \cos (\omega + \Omega + \phi)
\nonumber \\
& & + \ffrac{3}{4}a^2 \aplanet^2 \sin^4 \halfi 
\nonumber 
\end{eqnarray}
We are indebted to Seppo Mikkola, who kindly confirmed the correctness
of these expressions for us. The formula given in Mikkola et
al. (1994) is erroneous. Expansions to the fourth order in both
eccentricity and inclination are presented in Tabachnik (1999).

\end{appendix}

\end{document}